\documentclass{statsoc}
\usepackage{geometry}
\usepackage{graphicx}
\usepackage{amsmath}
\usepackage{natbib}
\usepackage{siunitx}
\usepackage[utf8]{inputenc} % allow utf-8 input
\usepackage{amsfonts}       % blackboard math symbols
\usepackage{nicefrac}       % compact symbols for 1/2, etc.
\usepackage{microtype}      % microtypography
\usepackage{amsmath, amsfonts,bm,amssymb,indentfirst,mathtools,bbm}
\usepackage[noend]{algpseudocode}
\usepackage{algorithmicx,algorithm,enumerate,accents}
\usepackage{color}
\usepackage{hyperref} 

\allowdisplaybreaks[4]

\numberwithin{table}{section}
\numberwithin{figure}{section}
\numberwithin{equation}{section}

\newtheorem{theorem}{Theorem}
\newtheorem{definition}{Definition}
\newtheorem{assumption}{Assumption}

\newtheorem{remark}{Remark}

\title[Estimation of the Epidemic Branching Factor in Noisy Contact Networks]{Estimation of the Epidemic Branching Factor in Noisy Contact Networks}

\author[Wenrui Li {\it et al.}]{Wenrui Li}
\address{Boston University,
	Boston,
	USA.}
\author{Daniel L. Sussman}
\address{Boston University,
	Boston,
	USA.}
\author{Eric D. Kolaczyk}
\coaddress{Eric D. Kolaczyk, Department of Mathematics \& Statistics, Boston University, 111 Cummington Mall, Boston MA, 02215, USA.\\ \sffamily Email: {kolaczyk@bu.edu}}
\address{Boston University,
	Boston,
	USA.}

\begin{document} 
	
\begin{abstract}
	Many fundamental concepts in network-based epidemic modeling depend on the branching factor, which captures a sense of dispersion in the network connectivity and quantifies the rate of spreading across the network.  Moreover, contact network information generally is available only up to some level of error. We study the propagation of such errors to the estimation of the branching factor. Specifically, we characterize the impact of network noise on the bias and variance of the observed branching factor for arbitrary true networks, with examples in sparse, dense, homogeneous and inhomogeneous networks. In addition, we propose a method-of-moments estimator for the true branching factor. We illustrate the practical performance of our estimator through simulation studies and with contact networks observed in British secondary schools and a French hospital.\\

 Keywords: Branching factor; Noisy network; Method-of-moments. 
\end{abstract}

\section{Introduction}

Epidemic modeling, while not at all new, has taken on renewed importance this year due to the COVID-19. Many key concepts in mathematical epidemiology depend on the branching factor -- for example, the basic reproduction number $R_0$. The latter is generally defined as the number of secondary infections expected in the early stages of an epidemic by a single infective in a population of susceptibles \citep{anderson1991infectious,diekmann2000mathematical}. The importance of $R_0$ in the study of epidemics arises from its role in so-called threshold theorems, which state under which conditions the presence of an infective individual in a population will lead to an epidemic \citep{whittle1955outcome}. In network-based susceptible-exposed-infectious-removed (SEIR) models, $R_0$ can be shown to equal $\theta(\kappa-1)/(\theta+\gamma)$.  Here $\theta$ and $\gamma$ are infection and recovery rates, respectively (\cite{trapman2016inferring}), while the branching factor, $\kappa$, is a measure of heterogeneity of a network. The branching factor captures a notion of the average degree of the vertex reached by following an edge from a vertex and, therefore, measures the rate of spreading across the network. It is evident that knowing the value of $\kappa$ is vital for effective control responses in the early stages of an epidemic. In addition, various thresholds in epidemiological and percolation theory rely on the branching factor. In the discussion section, we provide details on how knowledge of the branching factor informs those statistics.

Increasingly, contact networks are playing an important role in the study of epidemiology. Knowledge of the structure of the network allows models to take into account individual-level behavioral heterogeneities and shifts. Network-based approaches have been explored for investigating disease outbreaks in human (\cite{eubank2004modelling}), livestock (\cite{kao2006demographic}) and wildlife (\cite{craft2009distinguishing}) populations. Moreover, contact network information generally is available only up to some level of error -- also known as network noise. For example, there is often measurement error associated with network constructions, where, by `measurement error' we will mean true edges being observed as non-edges, and vice versa. Such edge noise occurs in self-reported contact networks where participants may not perceive and recall all contacts correctly (\cite{smieszek2012collecting}). It can also be found in sensor-based contact networks where automated proximity loggers are used to report frequency and duration of contacts (\cite{drewe2012performance}). Contact tracing, and the contact networks that result, currently is playing a central role in the fight to control COVID-19 globally (especially in conjunction with testing) (\cite{cevik2020sars}, \cite{juneau2020effective}, \cite{kretzschmar2020impact}). We investigate how network noise impacts on the observed value of $\kappa$ and, therefore, on our understanding of infectious diseases spreading. 

Extensive work regarding uncertainty quantification has been done in the field of non-network epidemic modeling, where populations are assumed uniform and with homogeneous mixing. Given adequate data, estimates of model parameters, such as $\theta$ and $\gamma$, can be produced with accompanying standard errors. Methods for this purpose are reviewed in \citet[Chapter~9--12]{andersson2012stochastic} and \cite{becker1999statistical}. Many studies have explored the effects of uncertainty in parameter estimation on basic epidemic quantities. For instance, there have been efforts to quantify uncertainty in $R_0$ around recent high profile emergent events, including severe acute respiratory syndrome (SARS) (\cite{chowell2004model}), the new influenza A (H1N1) (\cite{white2009estimation}), and Ebola (\cite{chowell2004basic}). But, to our best knowledge, there has been little attention to date given towards uncertainty analysis of $\kappa$ and relevant quantities in network-based epidemic models. Exceptions include real-time estimation of $R_0$ at an early stage of an outbreak by considering the heterogeneity in contact networks (\cite{davoudi2012early}), and measurability of $R_0$ in highly detailed sociodemographic data with the clustered contact structure assumed of the population (\cite{liu2018measurability}).

As remarked above, there appears to be little in the way of a formal and general treatment of the error propagation problem in network-based epidemic models. However, there are several areas in which the probabilistic or statistical treatment of uncertainty enters prominently in network analysis. Model-based approaches include statistical methodology for predicting network topology or attributes with models that explicitly include a component for network noise (\cite{jiang2011network}, \citet{jiang2012latent}), the `denoising' of noisy networks (\cite{chatterjee2015matrix}), the adaptation of methods for vertex classification using networks observed with errors (\cite{priebe2015statistical}), and a general Bayesian framework for reconstructing networks from observational data (\cite{young2020robust}).
The other common approach to network noise is based on a `signal plus noise' perspective. For example, \cite{balachandran2017propagation} introduced a simple model for noisy networks that, conditional on some true underlying network, assumed we observed a version of that network corrupted by an independent random noise that effectively flips the status of (non)edges. Later, \cite{chang2020estimation} developed method-of-moments estimators for the underlying rates of error when replicates of the observed network are available. In a somewhat different direction, uncertainty in network construction due to sampling has also been studied in some depth. See, for example, \citet[Chapter~5]{kolaczyk2009statistical} or \cite{ahmed2014network} for surveys of this area. However, in this setting, the uncertainty arises only from sampling---the subset of vertices and edges obtained through sampling are typically assumed to be observed without error.

Our contribution in this paper is to quantify how such errors propagate to the estimation of the branching factor, and to provide estimators for $\kappa$ when as few as three replicates of the observed network are available. Adopting the noise model proposed by \cite{balachandran2017propagation}, we characterize the impact of network noise on the bias and variance of the observed branching factor for arbitrary true networks, and we illustrate the asymptotic behaviors of these quantities on networks for varying densities and degree distributions. Our work shows that, in general, the bias in empirical branching factors can be expected to be nontrivial and is likely to dominate the variance.  Accordingly, we propose a parametric estimator of the branching factor, motivated by \cite{chang2020estimation}, who recently developed method-of-moments estimators for network subgraph densities and the underlying rates of error when replicates of the observed network are available. Numerical simulation suggests that high accuracy is possible for estimating branching factors in networks of even modest size. We illustrate the practical use of our estimators in the context of contact networks in British secondary schools and a French hospital, where a small number of replicates are available.

The organization of this paper is as follows. In Section \ref{sec2} we provide background on the noise model and branching factor. In Section \ref{sec3} we then present results for the bias and variance of the observed branching factor in sparse, dense, homogeneous and inhomogeneous networks. Section \ref{sec5} proposes our method-of-moments estimator for the true branching factor. Numerical illustration is reported in Section \ref{sec6}. All proofs are relegated to supplementary materials.

\section{Background}\label{sec2}

In this section, we provide essential notation and background.

\subsection{Noise model}

We assume the observed graph is a noisy version of a true graph. Let $G=(V,E)$ be an undirected graph and $G^\text{obs}=(V,E^\text{obs})$ be the observed graph, where we implicitly assume that the vertex set $V$ is known. Denote the adjacency matrix of $G$ by $\bm A=(A_{i,j})_{N_v\times N_v}$ and that of $G^\text{obs}$ by $\tilde {\bm A}=(\tilde A_{i,j})_{N_v\times N_v}$. Hence $A_{i,j} = 1$ if there is a true edge between the $i$-th vertex and the $j$-th vertex, and 0 otherwise, while $\tilde A_{i,j} = 1$ if an edge is observed between the $i$-th vertex and the $j$-th vertex, and 0 otherwise. And denote the degree of the $i$-th vertex in $G$ and $G^\text{obs}$ by $d_i$ and $\tilde d_i$, respectively. We assume throughout that $G$ and $G^\text{obs}$ are simple.

We express the marginal distributions of the $\tilde A_{i,j}$ in the form (\cite{balachandran2017propagation}): 
\begin{equation*}\label{eq2.1}
\begin{aligned}
\tilde A_{i,j}\sim
\begin{cases}
\text{Bernoulli}(\alpha_{i,j}), & \text{if } \{i,j\}\in E^c\\
\text{Bernoulli}(1-\beta_{i,j}), & \text{if } \{i,j\}\in E,\\
\end{cases}
\end{aligned}
\end{equation*}
where $E^c=\{\{i,j\} : i,j\in V; i< j\}  \backslash E$. Drawing by analogy on the example of network construction based on hypothesis testing, $\alpha_{i,j}$ can be interpreted as the probability of a Type-I error on the (non)edge status for vertex pair $\{i,j\}\in E^c$, while $\beta_{i,j}$ is interpreted as the probability of Type-II error, for vertex pair $\{i,j\}\in E$.

Our interest is in characterizing the manner in which the uncertainty in the $\tilde A_{i,j}$ (as a noisy version of $A_{ij}$) propagates to the branching factor. Here we focus on a general formulation of the problem in which we make the following three assumptions.

\begin{assumption}[Constant marginal error probabilities]\label{a1}
	Assume that \\ $\alpha_{i,j}=\alpha$ and $\beta_{i,j}=\beta$ for all $i< j$, so the marginal error probabilities are $\mathbb P(\tilde A_{i,j}=0|A_{i,j}=1)=\beta$ and $\mathbb P(\tilde A_{i,j}=1|A_{i,j}=0)=\alpha$.
\end{assumption}

\begin{assumption}[Independent noise]\label{a2}
	The random variables $\tilde A_{i,j}$, for all $i<  j$, are conditionally independent given $A_{i,j}$.
\end{assumption}

\begin{assumption}[Large Graphs]\label{a3}
	$N_v\rightarrow\infty$.
\end{assumption}

In Assumption \ref{a1}, we assume that both $\alpha$ and $\beta$ remain constant over different edges. Under Assumption \ref{a2}, the distributions of $\tilde d_i$ is
\begin{equation*} 
\begin{aligned}
\tilde d_i=\sum_{j=1}^{N_v} \tilde A_{j,i} \sim 
\text{Binomial}(N_v-1-d_i,\alpha ) + \text{Binomial}(d_i,1-\beta ).
\end{aligned}
\end{equation*}
Assumption \ref{a2} is not strictly necessary. See Remark \ref{remark3} in Section \ref{sec5}. Assumption \ref{a3} reflects both the fact that the study of large graphs is a hallmark of modern applied work in complex networks and, accordingly, our desire to understand the asymptotic behavior of the branching factor and provide concise descriptions in terms of the bias and variance for large graphs. 

\begin{remark}
	Note that $\alpha$ and $\beta$ can be constants or approach 0 as $N_v\rightarrow \infty$. For example, under Assumption \ref{a4}, if $\beta$ is constant and $|E|$ is dominated by $|E^c|$ asymptotically, then $\alpha$ approaches 0 as $N_v\rightarrow \infty$. Thus, $\alpha$ and $\beta$ are actually $\alpha(N_v)$ and $\beta(N_v)$. For notational simplicity, we omit $N_v$.
\end{remark}

In addition to the core Assumptions \ref{a1} -- \ref{a3}, we add a fourth assumption, upon which we will call periodically throughout the paper when desiring to illustrate our results in the special case.

\begin{assumption}[Edge Unbiasedness]\label{a4}
	$\alpha |E^c|=\beta |E|$, so that the expected number of observed edges equals the actual number of edges.
\end{assumption}

Our use of Assumption \ref{a4} reflects the understanding that a `good’ observation $G^\text{obs}$ of the graph $G$ should at the very least have roughly the right number of edges. 

\begin{remark}\label{r2}
	Assumption \ref{a4} cannot guarantee the unbiasedness of higher-order subgraph counts. (\cite{balachandran2017propagation})	
\end{remark}

\subsection{The branching factor in network-based epidemic models}

In general, the epidemic threshold of a network is the inverse of the largest eigenvalue of the adjacency matrix. Under some configuration models, the branching factor is often a good approximation of the largest eigenvalue (\cite{pastor2015epidemic}).

Let $G$ be a network graph describing the contact structure among $N_v$ elements in a population. If $G$ derives from a so-called configuration model, as is commonly assumed in the network-based epidemic modeling literature, then the branching factor takes the following form (\cite{buono2014epidemics}).

\begin{definition} 
	For graph $G$ with $N_v$ nodes, the branching factor is 
	\begin{align*}
	\kappa=\begin{cases}
	\displaystyle  \frac{\sum_{i=1}^{N_v} d_i^2/N_v}{\sum_{i=1}^{N_v} d_i/N_v} & \text{if }\sum_{i=1}^{N_v} d_i>0\\
	0&  \text{if }\sum_{i=1}^{N_v} d_i=0,
	\end{cases} 
	\end{align*}
	where $d_i$ is the degree of node $i$.
\end{definition}

Accordingly, we denote the branching factor in the noisy network by $\tilde \kappa$. Besides the basic reproduction number, $R_0$, described in the introduction, there are other quantities depending on the observed branching factor. These include the percolation threshold $1/(\tilde \kappa-1)$, the epidemic threshold $1/(\tilde \kappa-1)$, and the immunization threshold $1-1/(\lambda\tilde\kappa)$, where $\lambda$ is the spreading rate (\cite{pastor2015epidemic}).

\section{Bias and variance of the observed branching factor}\label{sec3}

In this section, we first quantify the asymptotic bias and variance of the observed branching factor for four typical classes of networks: sparse and homogeneous, sparse and inhomogeneous, dense and homogeneous, and dense and inhomogeneous.  We then provide numerical illustrations. In the supplementary material A and B, we present general results for the asymptotic bias and variance of the observed branching factor in arbitrary true networks. See supplementary material C - F for all proofs related to the observed branching factor.

\subsection{Bias of the observed branching factor}\label{sec3.2}

By making assumptions on the network density and degree distribution, we can obtain a nuanced understanding of the limiting behavior of the observed branching factor in terms of bias when the number of nodes tends towards infinity. Specifically, we consider the combinations of sparse versus dense and homogeneous versus inhomogeneous networks. By the term sparse we will mean a graph for which the average degree $\bar d$ is bounded both above and below by $\log N_v$ asymptotically, and by dense, $\bar d$ is bounded both above and below by $N_v^c$ asymptotically, where $0<c< 1$. By the term homogeneous we mean the degrees follow a Poisson distribution, and by inhomogeneous, the degrees follow a truncated Pareto distribution.

\begin{theorem}[Sparse and homogeneous, dense and homogeneous] \label{coro1}
	We define $Y=\sum_{i=1}^{N_v} \tilde d_i$. In the sparse homogeneous graph and dense homogeneous graph, under Assumption \ref{a1} - \ref{a4}, $\mathbb EY>0$, $\mathbb EY$ is bounded below by $N_v$ asymptotically, and $\beta$ is bounded, we have that the bias of $\tilde \kappa$ is dominated by $\kappa$ asymptotically.
\end{theorem}

\begin{theorem}[Sparse and inhomogeneous, dense and inhomogeneous] \label{coro2}
	In the sparse inhomogeneous graph and dense inhomogeneous graph, under the assumptions in Theorem \ref{coro1}, we have
	
	(i) if $0<\zeta\leq  2$, the bias of $\tilde \kappa$ is equal to $-\beta (2-\alpha-\beta)\kappa$ asymptotically,
	
	(ii) if $\zeta> 2$, the bias of $\tilde \kappa$ is equal to $-\beta (2-\alpha-\beta)\dfrac{\kappa}{ (\zeta-1)^2}$ asymptotically,
	where $\zeta$ is the shape parameter of the truncated Pareto distribution.
\end{theorem}

In summary, the observed branching factor is asymptotically unbiased in the homogeneous network setting, but asymptotically biased in the inhomogeneous network setting. The bias of the observed branching factor is negative which reflects the fact that the observed graph is typically more homogeneous then the true graph in the inhomogeneous setting. The bias depends on $\alpha$, $\beta$, and $\zeta$, and when the shape $\zeta>2$, the bias decreases as $\zeta$ increases. The different results in the homogeneous and inhomogeneous network setting also reflect Remark \ref{r2} since the branching factor is related to the second-order moment.

\subsection{Variance of the observed branching factor}

Again, by making assumptions on the network density and degree distribution, we can describe the limiting behavior of the observed branching factor in term of variance when the number of nodes tends towards infinity.

\begin{theorem}[Sparse, dense, homogeneous, and inhomogeneous]  
	In the combinations of sparse versus dense and  homogeneous versus inhomogeneous networks, under the assumptions in Theorem \ref{coro1}, we have that the variance of $\tilde \kappa$ is dominated by the bias of $\tilde \kappa$ asymptotically.
\end{theorem}

Note that the orders of the variances are asymptotically dominated by the corresponding biases for all four cases. Therefore, in noisy contact networks, bias would appear to be the primary concern for the observed branching factor.  In turn, our simulation results (below) suggest that in practice this empirical bias can be quite substantial.

\subsection{Simulation study}

\begin{figure}[!h]
	\centering
	\caption{Biases and variances of observed branching factors in homogeneous and inhomogeneous networks with different average degrees. Error bars are 95\% confidence intervals (and often not visible, due to the scale of bias versus variance). }	
	\includegraphics[width=5.5in, height=4in]{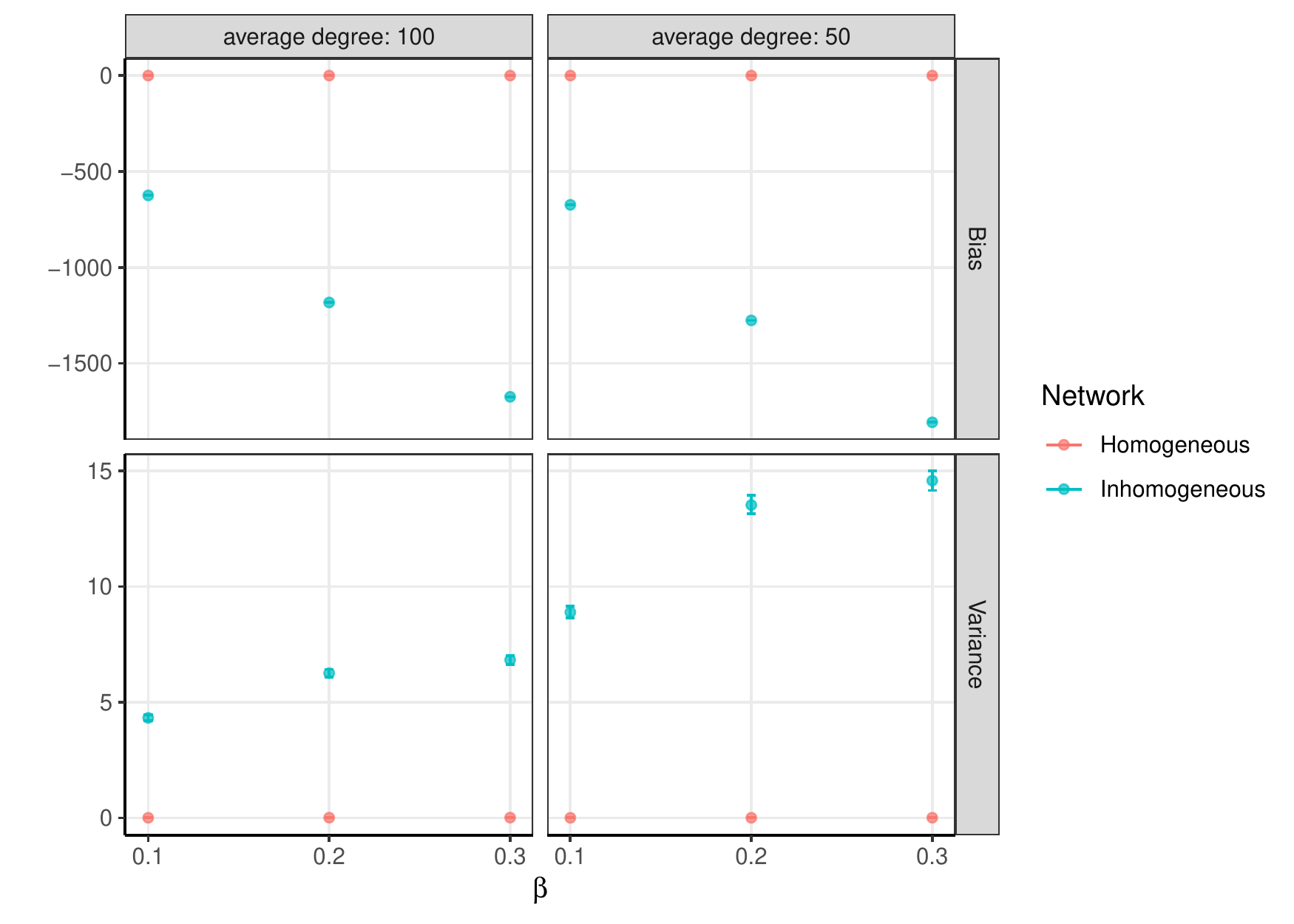}
	\label{fig0}
\end{figure}

We focus on two types of networks in the simulation study: random Erd\H{o}s-R\'{e}nyi networks and random scale-free networks using a preferential attachment mechanism.
The first type has a Poisson degree distribution, and the second type has a power law distribution. We construct two types of networks with 10,000 nodes and average degree around 50 or 100 and  view them as true networks. Then we generate 10,000 noisy, observed networks according to (\ref{eq2.1}). We set $\beta=0.1,0.2,0.3$ and $\alpha=\beta |E|/|E^c|$ (i.e., to enforce edge-unbiasedness). For each observed network, we compute $\tilde \kappa$. Also, we run 1,000 times bootstrap resampling to obtain 95\% confidence intervals for biases and variances. Biases and variances are shown in Figure \ref{fig0}. Error bars are 95\% confidence intervals.
 
 From the plots, we see that the noisy branching factor is unbiased in the homogeneous network setting, but biased in the inhomogeneous network setting. The bias of the observed branching factor is negative (i.e., the empirical branching factor generally underestimates the truth). And the bias increases when error rates increase. When the average degree increases from 50 to 100, the value of the true branching factor decreases from 3579.76 to 3356.34 and the bias decreases, which is consistent with Theorem \ref{coro2}. Also, variances are dominated by the corresponding biases in all cases.
 
\section{Estimator for the true branching factor}\label{sec5}

As we saw in Section \ref{sec3}, the observed branching factor is biased in the inhomogeneous network setting. Due to the presence of heterogeneity in the level of connectivity of contact neighborhoods for most real-world contact network data, it is important to have new estimators for bias reduction. Simultaneous estimation of Type I and II errors, $\alpha$ and $\beta$, as well as network quantities like $\kappa$, from a single noisy network is in general impossible \cite[Thm 1]{chang2020estimation}.  We present a method-of-moments estimator, which needs a minimum of three replicates.

We adapt the method-of-moments estimators (MME) of subgraph density in \cite{chang2020estimation}, which require at least three replicates of the observed network. Let $C_{\mathcal V_1}$ and $C_{\mathcal V_2}$ denote the edge density and the two-stars density, respectively. Then
\begin{align*}
C_{\mathcal V_1}=\frac{1}{|\mathcal V_1|}\sum_{\bm v=(i_1,i_1')\in\mathcal V_1} A_{i_1,i_1'} 
\end{align*}
and
\begin{align*}
C_{\mathcal V_2}=\frac{1}{|\mathcal V_2|}\sum_{\bm v=(i_1,i_1',i_2,i_2')\in\mathcal V_2} A_{i_1,i_1'} A_{i_2,i_2'}, 
\end{align*}
where $\mathcal V_1=\{(i_1,i_1'):i_1<i_1'\}$ and $\mathcal V_2=\{ (i_1,i_1',i_2,i_2') :i_1'=i_2, i_1\neq i_2\neq i_2' \}$.

Next we define 
\begin{align*}
\hat{\bar d}&=(N_v-1) \hat C_{\mathcal V_1},\\
\hat{\bar {d^2}}&= (N_v-1)(N_v-2)  \hat C_{\mathcal V_2}+ \hat{\bar d},\\
\end{align*}
where $\hat C_{\mathcal V_1}$ and $\hat C_{\mathcal V_2}$ are method-of-moments estimators of $C_{\mathcal V_1}$ and $C_{\mathcal V_2}$, which we will define later.
Thus, our estimator of $\kappa$ is given by: 
\begin{align}\label{eq5.1}
\hat \kappa= \frac{\hat{\bar {d^2}}}{\hat{\bar d}}= (N_v-2) \frac{\hat C_{\mathcal V_2}}{\hat C_{\mathcal V_1}}+1.
\end{align}

\begin{theorem}\label{thm:MME}
	Under Assumptions \ref{a1} and \ref{a2}, $\hat\kappa$ has asymptotic normal distribution with mean $\kappa$.
\end{theorem}

See supplementary material G for proof of Theorem \ref{thm:MME}. Note that $\hat \kappa$ is an asymptotically unbiased estimator for $\kappa$, where the asymptotics is in $N_v^2$, i.e., the square of the number of vertices in the network.  
To compute $\hat \kappa$, we first estimate $C_{\mathcal V_1}$ and $C_{\mathcal V_2}$ by methods used in \cite{chang2020estimation}. Define relevant quantities as follows:
\begin{align*}
u_1&= (1-\delta)\alpha+\delta(1-\beta),\\
u_2&= (1-\delta)\alpha(1-\alpha)+\delta\beta(1-\beta),\\
u_3&= (1-\delta)\alpha(1-\alpha)^2+\delta\beta^2(1-\beta),
\end{align*}
where $\delta$ is the edge density in the true network, $u_1$ is the expected edge density in one observed network, $u_2$ is the expected density of edge differences in two observed networks, and $u_3$ is the average probability of having an edge between two arbitrary nodes in one observed network but no edge between same nodes in the other two observed networks.  The method-of-moments estimators for $u_1$, $u_2$ and $u_3$ are
\begin{equation}\label{eq5.3}
\begin{aligned}
\hat u_1&=\frac{2}{N_v(N_v-1)}\sum_{i<j}\tilde A_{i,j}, \\
\hat u_2&=\frac{1}{N_v(N_v-1)}\sum_{i<j}|\tilde A_{i,j,*}-\tilde A_{i,j}|,\\
\hat u_3&=\frac{2}{3N_v(N_v-1)}\sum_{i<j} I( \text{Exactly one of }\tilde A_{i,j,**}, \tilde A_{i,j,*}, \tilde A_{i,j} \text{ equals } 1) .
\end{aligned}
\end{equation} 
where $ \tilde {\bm A}_*=(\tilde A_{i,j,*})_{N_v\times N_v},\ \tilde {\bm A}_{**}=(\tilde A_{i,j,**})_{N_v\times N_v}$ are independent and identically distributed replicates of $\tilde {\bm A}$. 

Calculation of the estimator $\hat\kappa$ in (\ref{eq5.1}) and the estimation of its asymptotic variance can be accomplished as detailed in Algorithm \ref{algo1} below and Algorithm 1 in the supplementary material H, respectively.  The variance estimation is based on a nonstandard bootstrap.

\begin{algorithm}[!h]
	\caption{Method-of-moments estimator $\hat \kappa$} 
	\hspace*{0.02in} {\bf Input:} 
	$\tilde {\bm A}=(\tilde A_{i,j})_{N_v\times N_v}, \ \tilde {\bm A}_*=(\tilde A_{i,j,*})_{N_v\times N_v},\ \tilde {\bm A}_{**}=(\tilde A_{i,j,**})_{N_v\times N_v},\ \alpha_0,\ \varepsilon$\\
	\hspace*{0.02in} {\bf Output:} 
	$\hat\alpha$, $\hat\beta$, $\hat \kappa$ 
	\begin{algorithmic}  
	   \State Compute $\hat u_1,\ \hat u_2,\  \hat u_3$ defined in (\ref{eq5.3});
		\State Initialize $\hat \alpha=\alpha_0$, $\alpha_0=\hat \alpha+10\varepsilon$;
		\While {$|\hat \alpha-\alpha_0|>\varepsilon$}
		\State $\alpha_0\gets\hat \alpha,\ \hat \beta\gets\frac{\hat u_2-\alpha_0+\hat u_1 \alpha_0}{\hat u_1-\alpha_0},\ \hat\delta\gets\frac{(\hat u_1-\alpha_0)^2}{\hat u_1-\hat u_2-2\hat u_1\alpha_0+\alpha_0^2},\ \hat\alpha\gets\frac{\hat u_3-\hat\delta\hat\beta^2(1-\hat\beta)}{(1-\hat\delta)(1-\alpha_0)^2}$;
		\EndWhile
		\State Compute $\hat k_3=1-\hat\alpha-\hat\beta,\ \hat C_{\mathcal V_1}=\frac{2}{\hat k_3N_v(N_v-1)}\sum_{i<j}(\tilde A_{i,j}-\hat\alpha)$,
        \State	\hspace{1.55cm} $\hat C_{\mathcal V_2}=\frac{1}{\hat k_3^2N_v(N_v-1)(N_v-2)}\sum_{i\neq j\neq l}(\tilde A_{i,j}-\hat\alpha)(\tilde A_{j,l}-\hat\alpha),\ \hat \kappa=(N_v-2) \frac{\hat C_{\mathcal V_2}}{\hat C_{\mathcal V_1}}+1$.
    \end{algorithmic}
	\label{algo1}
\end{algorithm}

\begin{remark}\label{remark3}
	Since our estimation of the unknown parameters is based on moment estimation, the independent noise dictated by Assumption \ref{a2} is not strictly necessary. As is shown in the proof of \cite{chang2020estimation}, the convergence rate for the moment estimation of the unknown parameters is determined by the convergence rates of $\hat u_1 -u_1$, $\hat u_2 -u_2$ and $\hat u_3 -u_3$. When some limited dependency among observed edges is present, the convergence rates of $\hat u_1 -u_1$, $\hat u_2 -u_2$ and $\hat u_3 -u_3$ still are bounded above by $1/N_v$ asymptotically.
\end{remark}

\section{Numerical illustration}\label{sec6}

In this section, we conduct some simulations and experiments to illustrate the finite sample properties of the proposed estimation methods. We consider two types of contact networks. One is the self-reported British secondary school contact network, described in \cite{kucharski2018structure}. These data were collected from 460 unique participants across four rounds of data collection conducted between January and June 2015 in year 7 groups in four UK secondary schools, with 7,315 identifiable contacts reported in total. They used a process of peer nomination as a method for data collection: students were asked, via the research questionnaire, to list the six other students in year 7 at their school that they spend the most time with. For each pair of participants in a specific round of data collection, a single link was defined if either one of the participants reported a contact between the pair (i.e. there was at least one unidirectional link, in either direction). Our analysis focuses on the single link contact network.

The other contact network we used is a sensor-based contact network in a French Hospital, reported by \cite{vanhems2013estimating}. These data contain records of contacts among patients and various types of health care workers in the geriatric unit of a hospital in Lyon, France, in 2010, from 1pm on Monday, December 6 to 2pm on Friday, December 10. Each of the 75 people in this study consented to wear RFID sensors on small identification badges during this period, which made it possible to record when any two of them were in face-to-face contact with each other (i.e., within 1-1.5 m of each other) during a 20-second interval of time.  A primary goal of this study was to gain insight into the pattern of contacts in such a hospital environment, particularly with an eye towards the manner in which infection might be transmitted. We define a link if duration of contacts in one day is greater than 5 minutes and construct networks for Tuesday, Wednesday and Thursday.

Each data set has at least three replicates.  And we consider two settings, a simulation setting where noise is added to a ‘true’ network derived from the data and an application setting where three replicates are each treated as noisy versions of an unknown true network.  The former results allow us to understand what finite-sample properties can be expected of our estimators, while the latter are reflective of what would be observed in practice with such data.

\subsection{Simulations}

For each data set, we artificially constructed a `true' adjacency matrix $\bm A$: if an edge occurs between a pair of vertices more than once in observed networks, we view that pair to have a true edge. The noisy, observed adjacency matrices $\tilde{\bm A}$, $\tilde{\bm A}_*$, $\tilde{\bm A}_{**}$ are generated according to (\ref{eq2.1}). We set $\alpha=0.005$ or 0.010, and $\beta=0.01,\ 0.15$, or 0.20. We assume that both $\alpha$ and $\beta$ are unknown. 

\begin{figure} 
	\centering
	\caption{Mean absolute errors (MAE) of $\hat \kappa$, and  95\% confidence intervals for $\kappa$ in the simulation with 500 replications for noisy networks in the hospital and schools. Reported in the plots are the relative frequencies (RF) of the event that a confidence interval covers the corresponding true value, and also the average Length of the intervals.}	
	\includegraphics[width=5.5in, height=7in]{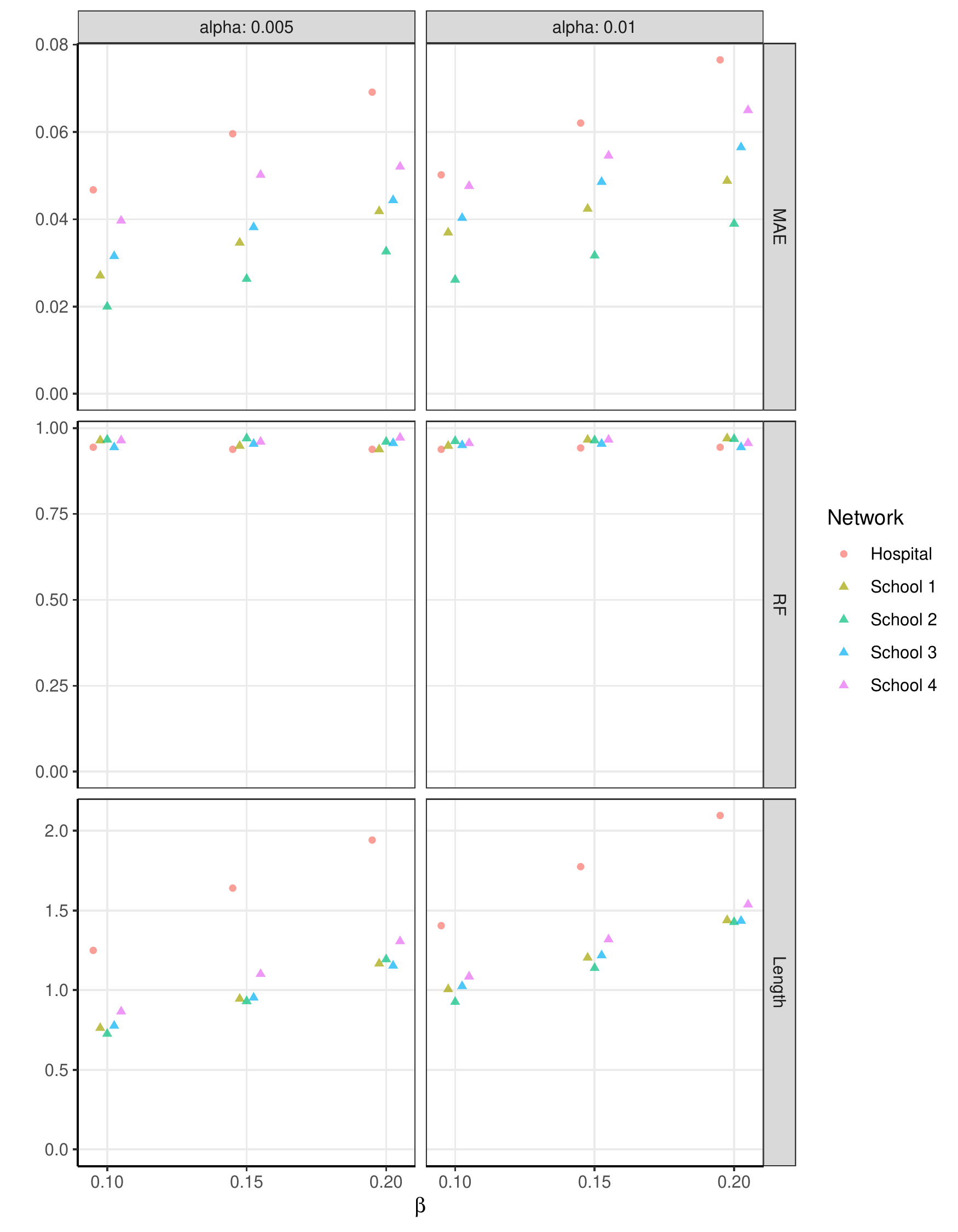}
	\label{fig2}
\end{figure}

We evaluate the method-of-moments estimate for $\kappa$ and 95\% confidence intervals. Figure \ref{fig2} shows the simulation results, in which we replicate 500 times for each setting. The mean absolute errors (MAE) for the point estimates for the branching factor $\kappa$ and the relative frequency (RF) of coverage for the estimated 95\% confidence interval for $\kappa$ are shown in Figure \ref{fig2}. Note that, $\text{MAE}(\hat\kappa)=\frac{1}{500}\sum_{i=1}^{500}|\hat\kappa_i-\kappa|$, where $\hat\kappa_1,\cdots,\hat\kappa_{500}$ denote the estimated values in 500 replications of simulation, and $\kappa$ denotes the true value.

In the hospital and school networks, the estimation errors for $\kappa$ increase when $\alpha$ and $\beta$ increase. And the estimated coverage probabilities are indeed around 95\%. The average interval lengths in the French hospital are larger than that in the four schools due to smaller sample size.

\subsection{Application}

In the school data sets, the nodes are not all the same within a given school over the four rounds. So, we choose the nodes common over four rounds and their edges to obtain four replicates of the noisy networks. Since our estimation methods only need three replicates, we select rounds 1, 2, and 3 (analogous results hold for other choices). Similarly, for the hospital data set, we choose the nodes common over three days and their edges to obtain three replicates of the noisy networks.

\begin{figure}[!h]  
	\centering
	\caption{The point estimates and 95\% confidence intervals for $\kappa$ in the hospital and four schools and the observed branching factor $\tilde \kappa$ in each round/day.}
	\includegraphics[width=4.8in, height=2.5in]{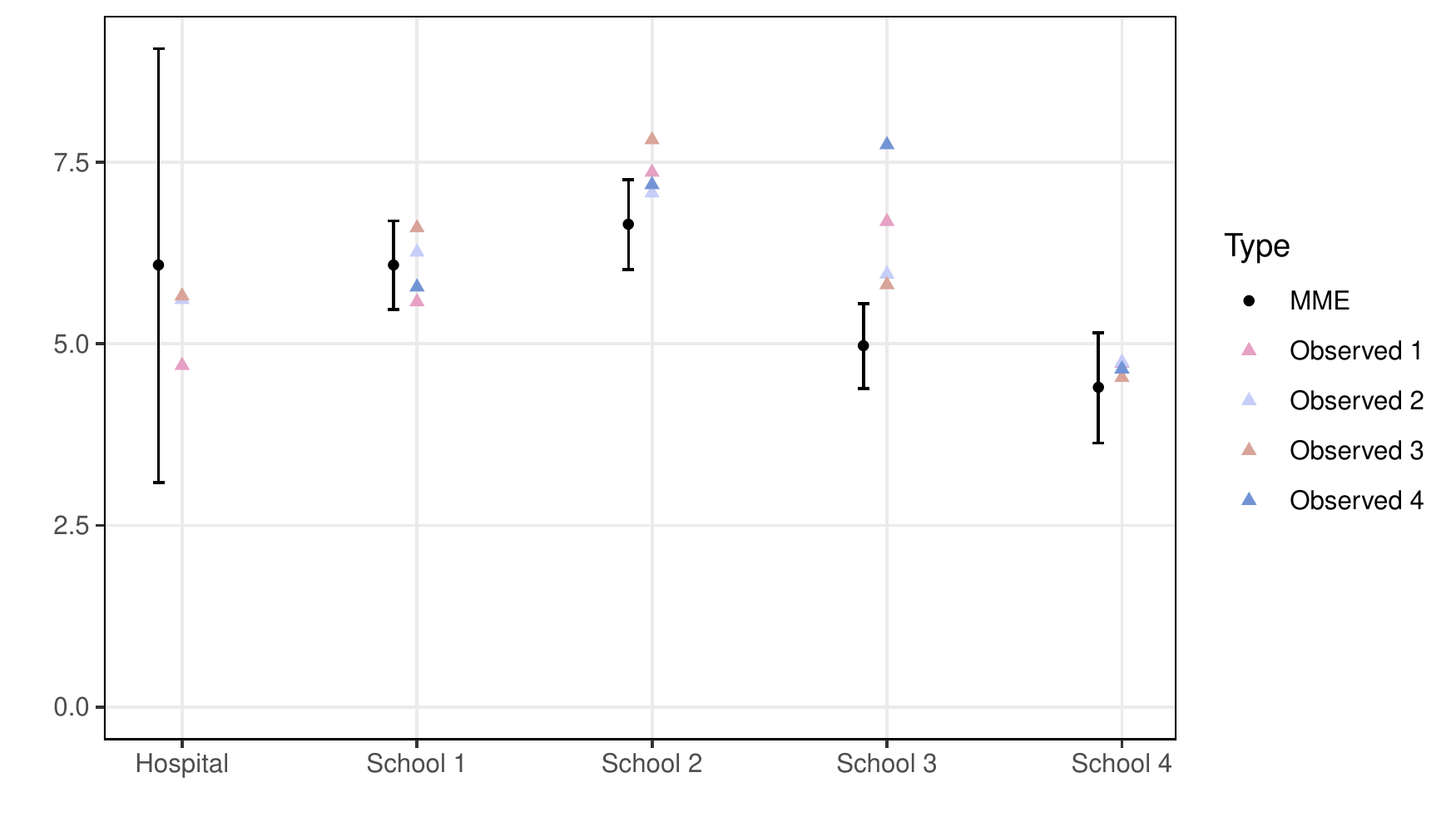}
	\label{fig1}
\end{figure}

We evaluate the method-of-moments estimates for $\kappa$, 95\% confidence intervals, and the observed branching factor $\tilde \kappa$. Point estimates and 95\% confidence intervals for $\alpha$ and $\beta$ are reported in Table \ref{table2}. Figure \ref{fig1} show the point estimates for the branching factor $\kappa$ and the observed branching factor $\tilde \kappa$ in each round. The error bars are the estimated 95\% confidence interval for $\kappa$.

Table \ref{table2} indicates there exists nontrivial noise in all networks.  The estimate of $\alpha$ in the hospital network is one order of magnitude larger than that in the school networks. Figure \ref{fig1} shows that, in schools 2 and 3, the resulting method-of-moments estimates for $\kappa$ are lower than all of their observed values, indicating a nontrivial downward adjustment for network noise. And most of the observed branching factors are not in the estimated 95\% confidence intervals, which further reinforces the evidence that the true branching factor is less than those observed empirically. In schools 1 and 4, the resulting method-of-moments estimates for $\kappa$ are close to their observed values. In contrast, in the French hospital, the estimate for $\kappa$ is higher than all of their observed values, indicating a nontrivial upward adjustment. 

Ultimately, we see that the ability to account for network noise appropriately in reporting the branching factor can lead to substantially different conclusions than use of the original, empirically observed branching factor.  These differences can then in turn be translated to specific epidemic-related quantities of interest in a study.

\begin{table}  
	\caption{Point estimates and 95\% confidence intervals for $\alpha$ and $\beta$ in the hospital and four schools. }
	% 	\centering
%	\begin{center}
	\fbox{ 
		\begin{tabular}{l@{\hskip  0.6cm} S[table-format = 1.1]
				@{\hskip  0.8cm(\,\hskip  -.3cm  }S[table-format = -1.2]@{ \hskip  .1cm,\,\hskip  -.3cm}S[table-format =  - 1.2]@{\,\hskip  .2cm) \ }
				@{\hskip  0.6cm} S[table-format = 1.1]
				@{\hskip  0.6cm(\,\hskip  -.1cm  }S[table-format = -1.2]@{ \hskip  .1cm,\,\hskip  -.1cm}S[table-format =  - 1.2]@{\,\hskip  .2cm) \ }
			}
			\\[-.8em] & \multicolumn{3}{c}{$\alpha$ } & \multicolumn{3}{c}{$\beta$ } \\
			\\[-.8em]
			Networks      & \multicolumn{1}{c@{\quad\space}}{Estimates} & \multicolumn{2}{c@{\hskip .8cm}}{CI}  & \multicolumn{1}{c@{\hskip 0.9cm}}{Estimates} & \multicolumn{2}{c}{CI} \\
			\hline
			\\[-.8em]
			Hospital & 0.116 & 0.080 &0.153 & 0.162 & -0.173& 0.499 \\ 
			School 1 &0.005 & 0.004 & 0.007 & 0.207 & 0.140 & 0.275 \\ 
			School 2 &0.013 & 0.012 & 0.015 & 0.141 & 0.092 & 0.191 \\  
			School 3 &0.013 & 0.012 & 0.015 & 0.000  & -0.057 & 0.057 \\ 
			School 4 &0.020 & 0.014 & 0.025 & 0.123 & 0.025 & 0.222 \\ 	
	\end{tabular}}
%	\end{center}
	\label{table2}
\end{table}

\begin{figure}[!h]
	\centering
	\caption{The point estimates and 95\% confidence intervals for $R_0$ in the hospital and four schools. }
	\includegraphics[width=4.8in, height=2.3in]{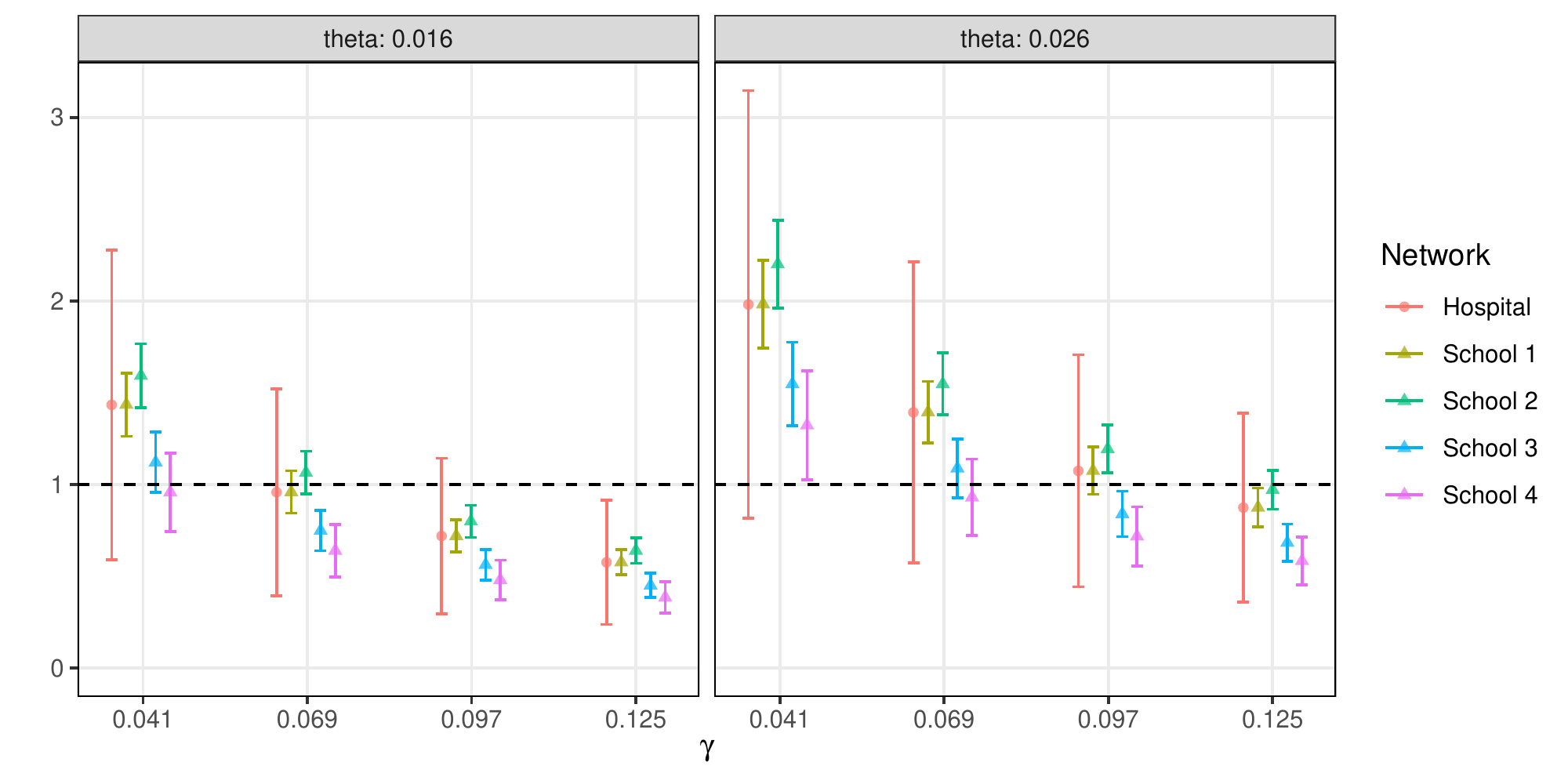}
	\label{fig3}
\end{figure}

For example, recall that $R_0$ equals $\theta(\kappa-1)/(\theta+\gamma)$ in the network-based SEIR model, where $\theta$ and $\gamma$ are infection and recovery rates. Therefore, if we are interested in characterizing the manner in which the uncertainty in the branching factor propagates to $R_0$, we can do so given knowledge or conjecture of values for these rates.  Consider the context of COVID-19, for example, for which current best knowledge suggests parameter settings of $\theta=0.016$ or $0.026$ and $1/\gamma$ from 8 to 24.6 (\cite{luo2020modes,lauer2020incubation,linton2020incubation,wang2020clinical,wolfel2020virological,verity2020estimates}). Estimating infection and recovery rates are important in epidemic modeling, but we treat $\theta$ and $\gamma$ as constants here for illustration, and only consider the uncertainty in the branching factor.

Figure \ref{fig3} shows the point estimates and 95\% confidence intervals for $R_0$ in the hospital and four schools. School 2 consistently has the highest estimated $\hat R_0$. The infection will be able to start spreading in a population when $R_0>1$, but not if $R_0<1$. For school networks, most of the 95\% confidence intervals include 1 or are below 1 when $\theta=0.016$, while some are higher when $\theta=0.026$. The 95\% confidence intervals include 1 in all cases for the French hospital.

\section{Discussion}

Here we have quantified the bias and variance of the observed branching factor in noisy networks and developed a general framework for estimation of the true branching factor in contexts wherein one has observations of noisy networks. Our approach requires as few as three replicates of network observations, and employs method-of-moments techniques to derive estimators and establish their asymptotic consistency and normality. Simulations demonstrate that substantial inferential accuracy by method-of-moments estimators is possible in networks of even modest size when nontrivial noise is present. And our application to contact networks in British secondary schools and a French hospital shows that the gains offered by our approach over presenting the observed branching factor can be pronounced.

We have pursued a frequentist approach to the problem of uncertainty quantification for the branching factor. If the replicates necessary for our approach are unavailable in a given setting, a Bayesian approach is a natural alternative. For example, posterior-predictive checks for goodness-of-fit based on examination of a handful of network summary measures is common practice (e.g., \cite{bloem2018random}). Note, however, that the Bayesian approach requires careful modeling of the generative process underlying $G$ and typically does not distinguish between signal and noise components. Our analysis is conditional on $G$, and hence does not require that $G$ be modeled. It is effectively a `signal plus noise’ model, with the signal taken to be fixed but unknown. Related work has been done in the context of graphon modeling, with the goal of estimating network motif frequencies (e.g., \cite{latouche2016variational}). However, again, one typically does not distinguish between signal and noise components in this setting. Additionally, we note that the problem of practical graphon estimation itself is still a developing area of research.

Our work here sets the stage for extensions to various thresholds and statistics which depend on the branching factor. Recall that these include the percolation threshold $1/(\kappa-1)$, the epidemic threshold $1/(\kappa-1)$, and the immunization threshold $1-1/(\lambda\kappa)$, where $\lambda$ is the spreading rate (\cite{pastor2015epidemic}). Replacing $\kappa$ with $\hat\kappa$, we obtain asymptotically unbiased estimators for the corresponding thresholds. The asymptotic distributions can be derived from the delta method. In addition, the total branching factor of the network is important for epidemic spreading and immunization strategy in multiplex networks (e.g., \cite{buono2014epidemics}).

Our choice to work with independent network noise is both natural and motivated by convenience. And our results of method-of-moments estimators still hold when there is some dependency across (non)edges. A precise characterization of the dependency is typically problem-specific and hence a topic for further investigation.

\section{Data accessibility} 

No primary data are used in this paper. Secondary data sources are taken from \cite{kucharski2018structure} and \cite{vanhems2013estimating}. These data and the code necessary to reproduce the results in this paper are available at \url{https://github.com/KolaczykResearch/EstimNetReprodNumber}.

\section{Acknowledgement}
 
This work was supported in part by ARO award W911NF1810237. This work was also supported by the Air Force Research Laboratory and DARPA under agreement number FA8750-18-2-0066 and by a grant from MIT Lincoln Labs.

%\bibliographystyle{rss}
%\bibliography{reference.bib}

\end{document}

\begin{abstract}
We provide theorems and corollaries for asymptotic bias and variance of the observed branching number and the proofs, proofs of theorems for method-of-moments estimator $\hat\kappa$, and the algorithm for estimation of asymptotic variance of $\hat\kappa$. 
\end{abstract}	

In this Supplementary Materials document, all results are asymptotic and we make use of asymptotic notations for conciseness. The meanings of the symbols are as follows. We say $f=\mathcal O(g)$ if $f$ is bounded above by $g$ asymptotically. If $f$ is bounded below by $g$ asymptotically, we write $f=\Omega (g)$. If we have $f=\mathcal O(g)$ and $f=\Omega (g)$, then we say $f=\Theta(g)$. And  $f\sim g$ represents that $f$ is equal to $g$ asymptotically. Lastly, $f=o(g)$ means that $f$ is dominated by $g$ asymptotically.

\appendix

\section{Theorems and corollaries for bias of the observed branching number}\label{sec:a}

In this section, we first quantify the asymptotic bias of the observed branching factor for arbitrary true networks. We then show specific results for four typical classes of networks: sparse and homogeneous, sparse and inhomogeneous, dense and homogeneous, and dense and inhomogeneous. Note that Theorem 1 in the main paper corresponds to Corollary \ref{coro1} and Corollary \ref{coro3},  and Theorem 2 in the main paper corresponds to Corollary \ref{coro2} and Corollary \ref{coro4}. 

\begin{theorem}\label{th1}
	We define $X=\sum_{i=1}^{N_v} {\tilde d_i}^2$, $Y=\sum_{i=1}^{N_v} \tilde d_i$ and we assume $\mathbb EY>0$, and $\mathbb EY=\Omega(N_v)\ ( N_v\rightarrow\infty)$. Then, under Assumption 2, for any $\eta>0$, we have	
	\begin{align*} 
	\text{Bias}[\tilde \kappa] = \frac{\mathbb EX}{\mathbb EY} -\kappa +\mathcal O\Big(   \frac{1}{(\mathbb EY)^{1/(2+\eta)} }  \frac{ \mathbb EX}{\mathbb EY} \Big)  \text{ as } N_v\rightarrow\infty.
	\end{align*}
\end{theorem}

\begin{remark}
	Theorem 1 reflects the fact that, under certain assumptions, $\mathbb EX/\mathbb EY$ is a good approximation of $\mathbb E(X/Y \cdot I_{\{Y>0\} } )$, i.e., $\mathbb E(\tilde \kappa)$.
\end{remark}

\begin{theorem}  \label{th2}
	Under assumptions in Theorem \ref{th1} and Assumption 1 and 4, for any $\eta>0$, we have	
	\begin{align*} 
	\text{Bias}[\tilde \kappa] =(2-\alpha-\beta) \Big[ \alpha(N_v-1)+\beta -(\alpha+\beta)\kappa \Big]+\mathcal O\Big(   \frac{1}{(\mathbb EY)^{1/(2+\eta)} }  \frac{ \mathbb EX}{\mathbb EY} \Big) 
	\end{align*}
	$\text{ as } N_v\rightarrow\infty$.
\end{theorem}

Theorem \ref{th1} shows the asymptotic bias of the observed branching factor in terms of the expectations of the first and second moments of the observed under Assumption 2. Theorem \ref{th2} relies on Assumptions 1 -- 4 and provides a more explicit expression for the leading term of the asymptotic bias in this special case.

\begin{corollary}[Sparse and homogeneous] \label{coro1}
	In the sparse homogeneous graph, where the average degree $\bar d=\Theta(\log N_v)$ and the asymptotic degree distribution is the Poisson distribution with mean $\bar d$, under the assumptions in Theorem \ref{th2} and $\beta=\mathcal O(1)\ ( N_v\rightarrow\infty)$, for any $\eta>0$, we have
	\begin{align*}
	\text{Bias}[\tilde \kappa]=\mathcal O\Big(\frac{\log N_v}{ (N_v \log N_v)^{1/(2+\eta)} }\Big) \text{ as } N_v\rightarrow\infty,
	\end{align*}
	where $\kappa=\Theta(\log N_v)$.
\end{corollary}

\begin{corollary}[Sparse and inhomogeneous] \label{coro2}
	In the sparse inhomogeneous graph where the average degree $\bar d=\Theta(\log N_v)$ and the asymptotic degree distribution is truncated Pareto distribution with shape $\zeta$, lower bound $d_L$ and upper bound $N_v-1$, under the assumptions in Theorem \ref{th2} and  $\beta=\mathcal O(1)\ ( N_v\rightarrow\infty)$, for any $\eta>0$, we have
	\begin{align*}
	\text{Bias}[\tilde \kappa]=\begin{cases}
	-\beta (2-\alpha-\beta)\kappa+ \mathcal O\Big( \max\Big\{ \log N_v, \dfrac{\kappa}{(N_v\log N_v)^{1/(2+\eta)}}\Big\}  \Big)  & \text{ if } 0<\zeta\leq  2 \\
	-\beta (2-\alpha-\beta)\dfrac{\kappa}{ (\zeta-1)^2}  +\mathcal O (1)  & \text{ if } \zeta> 2
	\end{cases}
	\end{align*}
	$\text{ as } N_v\rightarrow\infty,$ where 
	\begin{align*}
	\kappa=\begin{cases}
	\Theta(N_v ), &\text{ if } 0<\zeta<1 \\
	\Theta(N_v/\log N_v), &\text{ if } \zeta=1 \\
	\Theta(N_v^{2-\zeta}\cdot \log ^{\zeta-1}N_v), &\text{ if } 1<\zeta<2 \\
	\Theta(\log ^2N_v), &\text{ if } \zeta=2 \\
	\Theta(\log N_v), &\text{ if } \zeta>2. \\
	\end{cases}
	\end{align*}
\end{corollary}

\begin{remark}
	In Corollary \ref{coro2}, by the definition of expectation, $\zeta$, $d_L$, $\bar d$ and $N_v$ satisfy the equation
	\begin{align*}
	\bar d =\int_{d_L}^{N_v-1}x\cdot \frac{\zeta d_L^\zeta }{1- \Big(\dfrac{d_L}{N_v-1}\Big )^\zeta } x^{-(\zeta+1)}dx.
	\end{align*}
	Under the condition $\bar d=\Theta(\log N_v)$, the relationship among them can be simplified. Similar relationships also hold in Corollary \ref{coro4}, \ref{coro6} and \ref{coro8}.
\end{remark}

Note that the $\mathcal O$ term in Corollary \ref{coro2} is dominated by the corresponding $\kappa$ asymptotically, so $\text{Bias}(\tilde \kappa)=\Theta(\kappa)$, reflecting the challenges of estimating $\kappa$ in under heterogeneous degree distributions. In contrast, $\text{Bias}(\tilde \kappa) = o(\kappa)$ in Corollary~\ref{coro1}.
 
\begin{corollary}[Dense and homogeneous] \label{coro3}
	In the dense homogeneous graph where the average degree $\bar d=\Theta(N_v^c)$, $0<c< 1$, and the asymptotic degree distribution is the Poisson distribution with mean $\bar d$, under the assumptions in Theorem \ref{th2} and  $\beta=\mathcal O(1)\ ( N_v\rightarrow\infty)$, for any $\eta>0$, we have
	\begin{align*}
	\text{Bias}[\tilde \kappa]=\mathcal O\Big(N_v^{c-\frac{c+1}{2+\eta} }\Big) \text{ as } N_v\rightarrow\infty,
	\end{align*}
	where $\kappa=\Theta( N_v^c)$.
\end{corollary}

\begin{corollary}[Dense and inhomogeneous] \label{coro4}
	In the dense inhomogeneous graph where the average degree $\bar d=\Theta(N_v^c)$, $0<c<1$, and the asymptotic degree distribution is truncated Pareto distribution with shape $\zeta$, lower bound $d_L$ and upper bound $N_v-1$, under the assumptions in Theorem \ref{th2} and  $\beta=\mathcal O(1)\ ( N_v\rightarrow\infty)$,  for any $\eta>0$, we have
	\begin{align*}
	\text{Bias}[\tilde \kappa]=\begin{cases}
	-\beta (2-\alpha-\beta)\kappa+\mathcal O\Big(\max\Big\{N_v^c,\dfrac{\kappa}{N_v^{(c+1)/(2+\eta)}}\Big \}   \Big) 
	& \text{ if } 0<\zeta\leq  2 \\
	-\beta (2-\alpha-\beta)\dfrac{\kappa}{ (\zeta-1)^2}  +\mathcal O (\max\{N_v^{2c-1},1\})   & \text{ if } \zeta> 2
	\end{cases}
	\end{align*}
	$\text{ as } N_v\rightarrow\infty,$ where 
	\begin{align*}
	\kappa=\begin{cases}
	\Theta(N_v ), &\text{ if } 0<\zeta<1 \\
	\Theta(N_v/\log N_v), &\text{ if } \zeta=1 \\
	\Theta(N_v^{2-\zeta+c(\zeta-1)} , &\text{ if } 1<\zeta<2 \\
	\Theta(N_v^c\cdot\log N_v), &\text{ if } \zeta=2 \\
	\Theta(N_v^c), &\text{ if } \zeta>2. \\
	\end{cases}
	\end{align*}
	
\end{corollary}

Note that the $\mathcal O$ term in Corollary \ref{coro4} is dominated by the corresponding $\kappa$ asymptotically, so $\text{Bias}(\tilde \kappa)=\Theta(\kappa)$.  In contrast, $\text{Bias}(\tilde \kappa) = o(\kappa)$ in Corollary~\ref{coro3}.

\section{Theorems and corollaries for variance of the observed branching number}

In this section, we first compute the asymptotic variance of the observed branching factor for arbitrary true networks. We then show specific results for the same four types of networks as in Section \ref{sec:a}. Note that Theorem 3 in the main paper corresponds to Corollary \ref{coro5} - \ref{coro8}.

\begin{theorem}\label{th3}
	We assume $\mathbb EY>0$, and $\mathbb EY=\Omega(N_v)\ ( N_v\rightarrow\infty)$. Then, under Assumption 2,
	
	(i) 
	\begin{align*}
	\text{Var}[\tilde \kappa]= \mathcal O\Bigg( \max\Big\{  \mathbb E\Big[\frac{(X\mathbb EY-Y\mathbb E X)^2}{(\mathbb E Y)^4} \Big],  \mathbb P(Y=0)\cdot\Big[\frac{ \mathbb EX}{\mathbb EY}   \Big]^2\Big\}\Bigg )
	\end{align*}
	as $N_v\rightarrow\infty$.
	
	(ii) For any $\eta,\lambda>0$, 
	\begin{align*}
	\text{Var}[\tilde \kappa]
	=&\ \mathbb E\Big[\frac{(X\mathbb EY-Y\mathbb E X)^2}{(\mathbb E Y)^4} \Big] + \mathcal O\Bigg( \max\Big\{  (\mathbb EY)^{-1/(2+\eta)}\cdot\mathbb E\Big[\frac{(X\mathbb EY-Y\mathbb E X)^2}{(\mathbb E Y)^4} \Big], \\
	& (\mathbb EY)^{-2/(2+\lambda)}\cdot\Big[\frac{ \mathbb EX}{\mathbb EY}  \Big]^2,\mathbb P(Y=0)\cdot\Big[\frac{ \mathbb EX}{\mathbb EY}  \Big]^2 \Big\}\Bigg )
	\end{align*}
	as $N_v\rightarrow\infty$.
\end{theorem}

\begin{theorem}\label{th4}
	Under the assumptions in Theorem \ref{th3}, Assumption 1 and 4, and $1-\beta=\Omega(N_v)\ (N_v\rightarrow \infty)$, 
	
	(i) 
	\begin{align*}
	\text{Var}[\tilde \kappa]=  \mathcal O\Bigg(\mathbb E\Big[\frac{(X\mathbb EY-Y\mathbb E X)^2}{(\mathbb E Y)^4} \Big]\Bigg )   \text{ as } N_v\rightarrow\infty.
	\end{align*}

	(ii) For any $\eta,\kappa>0$, 
	\begin{align*}
	\text{Var}[\tilde \kappa]=&\ \mathbb E\Big[\frac{(X\mathbb EY-Y\mathbb E X)^2}{(\mathbb E Y)^4} \Big] \\
	&+ \mathcal O\Bigg( \max\Big\{  (\mathbb EY)^{-1/(2+\eta)}\cdot\mathbb E\Big[\frac{(X\mathbb EY-Y\mathbb E X)^2}{(\mathbb E Y)^4} \Big],  (\mathbb EY)^{-2/(2+\lambda)}\cdot\Big[\frac{ \mathbb EX}{\mathbb EY}  \Big]^2 \Big\}\Bigg )
	\end{align*}
	as $N_v\rightarrow\infty$.
\end{theorem}

Theorem \ref{th3} (i) and Theorem \ref{th4} (i) provide upper bounds for variances of the observed branching factors. And Theorem \ref{th3} (ii) and Theorem \ref{th4} (ii) derive good approximations of variances if the $\mathcal O$ terms are dominated by the corresponding first terms asymptotically.

\begin{corollary}[Sparse and homogeneous]  \label{coro5}
	In the sparse homogeneous graph where the average degree $\bar d=\Theta(\log N_v)$ and the asymptotic degree distribution is the Poisson distribution with mean $\bar d$, under the assumptions in Theorem \ref{th2} and $\beta=\mathcal O(1)\ ( N_v\rightarrow\infty)$, we have
	\begin{align*}
	\text{Var}[\tilde \kappa]= \mathcal{O}\Big( \Big( \frac{\log N_v}{N_v}\Big)^{1/2}\Big)\text{ as } N_v\rightarrow\infty.
	\end{align*}
\end{corollary}

\begin{corollary}[Sparse and inhomogeneous]  \label{coro6}
	In the sparse inhomogeneous graph where the average degree $\bar d=\Theta(\log N_v)$ and the asymptotic degree distribution is truncated Pareto distribution with shape $\zeta$, lower bound $d_L$ and upper bound $N_v-1$, under the assumptions in Theorem \ref{th2} and $\beta=\mathcal O(1)\ ( N_v\rightarrow\infty)$, we have
	\begin{align*}
	\text{Var}[\tilde  \kappa]= \begin{cases}
	\mathcal{O} ( N_v/\log N_v ), & 0<\zeta<1\\
	\mathcal{O}( N_v/\log^2 N_v ), &\zeta=1\\
	\mathcal{O}( (N_v/\log N_v )^{2-\zeta }), &1<\zeta<5/2\\
	\mathcal{O}( (\log N_v/ N_v )^{1/2}), &\zeta\geq 5/2\\
	\end{cases}
	\end{align*}
	as $N_v\rightarrow\infty.$	
\end{corollary}

\begin{corollary}[Dense and homogeneous]  \label{coro7}
	In the dense homogeneous graph where the average degree $\bar d=\Theta( N_v^c)$, $0<c<1$, and the asymptotic degree distribution is the Poisson distribution with mean $\bar d$, under the assumptions in Theorem \ref{th2} and $\beta=\mathcal O(1)\ ( N_v\rightarrow\infty)$, we have
	\begin{align*}
	\text{Var}[\tilde  \kappa]= \mathcal{O}(N_v^{(c-1)/2})\text{ as } N_v\rightarrow\infty.
	\end{align*}
\end{corollary}

\begin{corollary}[Dense and inhomogeneous]   \label{coro8}
	In the dense inhomogeneous graph where the average degree $\bar d=\Theta( N_v^c)$, $0<c<1$, and the asymptotic degree distribution is truncated Pareto distribution with shape $\zeta$, lower bound $d_L$ and upper bound $N_v-1$, under the assumptions in Theorem \ref{th2} and $\beta=\mathcal O(1)\ ( N_v\rightarrow\infty)$, we have
	\begin{align*}
	\text{Var}[\tilde  \kappa]= \begin{cases}
	\mathcal{O} ( N_v^{1-c} ), & 0<\zeta<1\\
	\mathcal{O}( N_v^{1-c} /\log N_v ), &\zeta=1\\
	\mathcal{O}( N_v^{(2-\zeta)(1-c) }), &1<\zeta<5/2\\
	\mathcal{O}( N_v^{(c-1)/2 } ), &\zeta\geq 5/2\\
	\end{cases}
	\end{align*}
	as $N_v\rightarrow\infty.$	
\end{corollary}

Note that the orders of the variances are asymptotically dominated by the corresponding biases for all four cases. Therefore, in noisy contact networks, bias would appear to be the primary concern for the observed branching factor. The $\mathcal O$ notations for variances in the homogeneous networks are bounded above by those in the inhomogeneous networks of the same network density.

\section{Proofs of theorems for bias of the observed branching number}

\subsection{Proof of Theorem 1}
	
Recall $X = \sum_{i} \tilde{d}_i^2$ and $Y = \sum_i \tilde{d}_i$.
Note that 	
\begin{align*} 
\text{Bias}[\tilde \kappa] = \frac{\mathbb EX}{\mathbb EY} -\kappa +\mathcal O\Big(   \frac{1}{(\mathbb EY)^{1/(2+\eta)} }  \frac{ \mathbb EX}{\mathbb EY} \Big) 
\end{align*}
is equivalent to
\begin{align}\label{eq10.1}
\mathbb E[\tilde \kappa]-   \frac{ \mathbb EX }{\mathbb EY}  =  \mathcal O\Big(   \frac{1}{(\mathbb EY)^{1/(2+\eta)} }  \frac{ \mathbb EX}{\mathbb EY} \Big) .
\end{align}
By Jensen's inequality, we have
\begin{align} \label{eq10.2}
\Bigg| \mathbb E[\tilde \kappa]-\frac{ \mathbb EX}{\mathbb EY}  \Bigg|= \frac{1}{\mathbb EY} \Bigg| \mathbb E\Big[  \frac{ X(\mathbb EY-Y) }{Y  }\cdot I_{ \{ Y>0 \}} \Big]\Bigg| \leq  \frac{1}{\mathbb EY}\cdot\mathbb E\Big[  \frac{ X|\mathbb EY-Y|}{Y  }\cdot I_{ \{ Y>0 \}}\Big].
\end{align}	
Then by additivity of expectation, for $0<\delta< 1$, $\mathbb E\Big[  \frac{ X|\mathbb EY-Y|}{Y  }\cdot I_{ \{ Y>0 \}}\Big]$ in (\ref{eq10.2}) equals
\begin{align} \label{eq10.3}
\mathbb E\Big[  \frac{ X|\mathbb EY-Y|}{Y  }\cdot I_{ \{ Y>0 \}}\cdot I_{  \{|Y-\mathbb E Y|\geq\delta\mathbb E Y\} }  \Big]+\mathbb E\Big[  \frac{ X|\mathbb EY-Y|}{Y  }\cdot I_{ \{ Y>0 \}}\cdot I_{  \{|Y-\mathbb E Y|<\delta\mathbb E Y\} }  \Big].
\end{align}	

Next, we find the upper bounds of two terms in (\ref{eq10.3}). For the first term, by definitions of $X$ and $Y$, $X/Y\cdot I_{ \{ Y>0 \}}<N_v$ and $|\mathbb EY-Y|<N_v^2$. Thus, we have
\begin{align*}
\mathbb E\Big[  \frac{ X|\mathbb EY-Y|}{Y  }\cdot I_{ \{ Y>0 \}}\cdot I_{  \{|Y-\mathbb E Y|\geq\delta\mathbb E Y\} }  \Big]
< N_v^3\cdot \text{Pr}(  |Y-\mathbb E Y|\geq\delta\mathbb E Y ).
\end{align*}
Then, by Chernoff Bound, we obtain
\begin{align*}
\mathbb E\Big[  \frac{ X|\mathbb EY-Y|}{Y  }\cdot I_{ \{ Y>0 \}}\cdot I_{  \{|Y-\mathbb E Y|\geq\delta\mathbb E Y\} }  \Big]
<  2N_v^3\cdot \exp\Big(-\frac{\delta^2 \cdot\mathbb{E}Y}{6}\Big).
\end{align*}
For the second term, when $|Y-\mathbb E Y|<\delta\mathbb E Y$, $Y> (1-\delta)\mathbb E Y$. So, we obtain
\begin{align*}
\mathbb E\Big[  \frac{ X|\mathbb EY-Y|}{Y  }\cdot I_{ \{ Y>0 \}}\cdot I_{  \{|Y-\mathbb E Y|<\delta\mathbb E Y\} }  \Big] 
<  \frac{\delta}{(1-\delta) }\cdot \mathbb EX.
\end{align*}
By (\ref{eq10.2}), we show
\begin{align} \label{eq10.4}
\Bigg| \mathbb E[\tilde \kappa]-\frac{ \mathbb EX}{\mathbb EY}  \Bigg|  \leq  \frac{ 2N_v^3}{\mathbb EY}\cdot \exp\Big(-\frac{\delta^2 \cdot\mathbb{E}Y}{6}\Big)+\frac{\delta}{(1-\delta) }\cdot \frac{ \mathbb EX}{\mathbb EY}.
\end{align}	

Let $L_1(N_v)$ and $L_2(N_v)$ denote two terms on the right sides in (\ref{eq10.4}). We choose $\delta=( \mathbb EY)^{-1/(2+\eta)}$, $\eta>0$, such that
\begin{align*} 
L_1(N_v) =o\Big(  \frac{\mathbb EX}{\mathbb EY}\Big) \text{ and }L_2(N_v)=o\Big(  \frac{\mathbb EX}{\mathbb EY}\Big) \text{ as } N_v\rightarrow \infty,
\end{align*}
under the assumption $\mathbb EY=\Omega(N_v)$ as $N_v\rightarrow\infty$, $1-\delta=\mathcal O(1)$. By L'Hopital's rule, we have 
\begin{align*}
L_1(N_v)=o(L_2(N_v))\text{ as }N_v\rightarrow\infty.
\end{align*}
These imply (\ref{eq10.1}). 

\subsection{Proof of Theorem 2}

We compute $\mathbb EY$ and $\mathbb EX$ under Assumption 1 and 2,
\begin{align*}\label{eq9.5}
\mathbb EY&= \sum_{i=1}^{N_v} \mathbb E[\tilde d_i]=\sum_{i=1}^{N_v} \alpha(N_v-1-d_i)+(1-\beta)d_i\\
&= \alpha N_v(N_v-1)+(1-\alpha-\beta)\sum_{i=1}^{N_v} d_i, \text{ and }\\
\mathbb EX&=  \sum_{i=1}^{N_v} \mathbb E[{\tilde d_i}^2]= \sum_{i=1}^{N_v} \big(\text{var}[\tilde d_i] + (\mathbb E[\tilde d_i])^2\big)\\
&= \sum_{i=1}^{N_v}\Big( \alpha(1-\alpha)(N_v-1-d_i)\\
& \hskip .5cm +\beta(1-\beta)d_i + \big[ \alpha(N_v-1-d_i)+(1-\beta)d_i\big]^2\Big)\\
&= (1-\alpha-\beta)^2 \sum_{i=1}^{N_v} d_i^2+ \big[\beta(1-\beta)-\alpha(1-\alpha)\\
& \hskip .5cm + 2\alpha(N_v-1)(1-\alpha-\beta)\big]  \sum_{i=1}^{N_v} d_i + \alpha N_v(N_v-1)\big[ 1-\alpha+ \alpha(N_v-1)\big].\numberthis
\end{align*}	
Then, under Assumption 4, (\ref{eq9.5}) leads to
\begin{align*}
\mathbb EY&=\sum_{i=1}^{N_v} d_i,\\
\mathbb EX&=(1-\alpha-\beta)^2 \sum_{i=1}^{N_v} d_i^2+ (2-\alpha-\beta)\big[\alpha(N_v-1)+\beta\big]  \sum_{i=1}^{N_v} d_i.
\end{align*}	
Plugging the value of $\mathbb EX/\mathbb EY$ into the bias expression in Theorem \ref{th1} completes the proof. 

\section{Proofs of corollaries for bias of the observed branching number}
\subsection{Proof of Corollary 1}
	
	By homogeneity, we obtain
	\begin{align*}
	\sum_{i=1}^{N_v}d_i^2=(\bar d +1 )\bar dN_v.
	\end{align*}
	By edge unbiasedness, we have 
	\begin{align}
	\alpha=\frac{\beta  \bar d}{N_v-1-\bar d}.
	\end{align}
	Thus, 
	\begin{align}
	\alpha(N_v-1) +\beta -(\alpha+\beta) \frac{\sum_{i=1}^{N_v}d_i^2}{\sum_{i=1}^{N_v}d_i}=-\alpha,
	\end{align}
	and 
	\begin{align*}
	\frac{\mathbb E X}{\mathbb E Y}=\bar d +1 -\alpha (2-\alpha-\beta).
	\end{align*}
	By Theorem 2, for any $\eta>0$, we have  
	\begin{align*}
	\text{Bias}[\tilde \kappa]=\mathcal O\Big(\frac{\log N_v}{ (N_v \log N_v)^{1/(2+\eta)} }\Big) \text{ as } N_v\rightarrow\infty.
	\end{align*}

\subsection{Proof of Corollary 2}

First, we compute the first and second moments of truncated Pareto distribution.
\begin{align*} \label{eq10}
\int_{d_L}^{N_v-1}x\cdot \frac{\zeta d_L^\zeta }{1- \Big(\dfrac{d_L}{N_v-1}\Big )^\zeta } x^{-(\zeta+1)}dx
&=\begin{cases}
\zeta d_L^\zeta \cdot  \dfrac{\log\Big(\dfrac{N_v-1}{d_L}\Big ) } {1- \Big(\dfrac{d_L}{N_v-1}\Big )^\zeta } ,   \text{ if }  \zeta=1\\
\dfrac{\zeta d_L}{\zeta-1}\cdot  \dfrac{1- \Big(\dfrac{d_L}{N_v-1}\Big )^{\zeta-1} } {1- \Big(\dfrac{d_L}{N_v-1}\Big )^\zeta } ,   \text{ otherwise, }\\
\end{cases}\\
\int_{d_L}^{N_v-1}x^2\cdot \frac{\zeta d_L^\zeta }{1- \Big(\dfrac{d_L}{N_v-1}\Big )^\zeta } x^{-(\zeta+1)}dx
&=\begin{cases}
\zeta d_L^\zeta \cdot  \dfrac{\log\Big(\dfrac{N_v-1}{d_L}\Big ) } {1- \Big(\dfrac{d_L}{N_v-1}\Big )^\zeta } ,   \text{ if }  \zeta=2\\
\dfrac{\zeta d_L^2}{\zeta-2}\cdot  \dfrac{1- \Big(\dfrac{d_L}{N_v-1}\Big )^{\zeta-2} } {1- \Big(\dfrac{d_L}{N_v-1}\Big )^\zeta } ,   \text{ otherwise. }  \\
\end{cases}\numberthis
\end{align*}
Note that $\bar d =\sum_{i=1}^{N_v}d_i/N_v=\Theta(\log N_v)$. So, as $N_v\rightarrow\infty$, we obtain
\begin{align*}\label{eq23}
d_L\sim\begin{cases}
\Big( \dfrac{1-\zeta}{\zeta} \cdot\dfrac{\bar d}{N_v^{1-\zeta}} \Big)^{1/\zeta}, &\text{ if } 0<\zeta<1 \\
\bar d/\log N_v, &\text{ if } \zeta=1 \\
\dfrac{\zeta-1}{\zeta} \bar d, &\text{ if } \zeta>1 .
\end{cases}\numberthis
\end{align*}
Thus,  as $N_v\rightarrow\infty$, we have
\begin{align*}\label{eq24}
\frac{\sum_{i=1}^{N_v}d_i^2}{\sum_{i=1}^{N_v}d_i}\sim\begin{cases}
\dfrac{1-\zeta}{2-\zeta}N_v, &\text{ if } 0<\zeta<1 \\
N_v/\log N_v, &\text{ if } \zeta=1 \\
\dfrac{\zeta-1}{2-\zeta} \Big(\dfrac{\zeta-1}{\zeta} \Big)^{\zeta-1}N_v^{2-\zeta}\cdot (\bar d  )^{\zeta-1}, &\text{ if } 1<\zeta<2 \\
\bar d \cdot \log N_v/2, &\text{ if } \zeta=2 \\
\dfrac{(\zeta-1)^2}{\zeta(\zeta-2)} \bar d, &\text{ if } \zeta>2. \\
\end{cases}\numberthis
\end{align*}
By edge unbiasedness, we have 
\begin{align*}
\alpha=\frac{\beta  \bar d}{N_v-1-\bar d}=\Theta\Big(\frac{\log N_v}{N_v}\Big).
\end{align*}

(i) $0<\zeta\leq 2$

Note that
\begin{align*}
\alpha(N_v-1) +\beta -(\alpha+\beta) \frac{\sum_{i=1}^{N_v}d_i^2}{\sum_{i=1}^{N_v}d_i}
= -  \beta \kappa+\alpha(N_v-\kappa-1) +\beta  ,
\end{align*}
and 
\begin{align*}
\frac{\mathbb E X}{\mathbb E Y}=\Theta\Big( \frac{\sum_{i=1}^{N_v}d_i^2}{\sum_{i=1}^{N_v}d_i} \Big).
\end{align*}
By Theorem 2, for any $\eta>0$, we have  
\begin{align*}
\text{Bias}[\tilde \kappa]&=-\beta (2-\alpha-\beta)\kappa+ (2-\alpha-\beta) \big[\alpha(N_v-\kappa-1) +\beta   \big]+\mathcal O\Big(\dfrac{1}{(\mathbb E Y)^{1/(2+\eta)}}\dfrac{\mathbb E X}{\mathbb E Y} \Big) \\
&=-\beta (2-\alpha-\beta)\kappa+\mathcal O\Big(\max\Big\{\log N_v,\dfrac{\kappa}{(N_v\log N_v)^{1/(2+\eta)}}\Big \}   \Big) \end{align*}
$\text{ as } N_v\rightarrow\infty.$

(ii) $\zeta> 2$

Note that
\begin{align*}
\alpha(N_v-1) +\beta -(\alpha+\beta) \frac{\sum_{i=1}^{N_v}d_i^2}{\sum_{i=1}^{N_v}d_i}\sim-\dfrac{\beta \bar d}{\zeta(\zeta-2)} -\dfrac{\beta (\bar d)^2}{\zeta(\zeta-2)(N_v-1-\bar d)} +\beta,
\end{align*}
and 
\begin{align*}
\frac{\mathbb E X}{\mathbb E Y}=\mathcal O( \bar d).
\end{align*}
By Theorem 2, for any $\eta>0$, we have  
\begin{align*}
\text{Bias}[\tilde \kappa]&=-  (2-\alpha-\beta)\dfrac{\beta \bar d}{\zeta(\zeta-2)}  +\mathcal O (1) \\
&=-\beta (2-\alpha-\beta)\dfrac{\kappa}{ (\zeta-1)^2}  +\mathcal O (1) 
\end{align*}
$\text{ as } N_v\rightarrow\infty.$

\subsection{Proof of Corollary 3}

By homogeneity, we obtain
\begin{align*}
\sum_{i=1}^{N_v}d_i^2=(\bar d +1 )\bar dN_v.
\end{align*}
By edge unbiasedness, we have 
\begin{align}
\alpha=\frac{\beta  \bar d}{N_v-1-\bar d}.
\end{align}
Thus, 
\begin{align}
\alpha(N_v-1) +\beta -(\alpha+\beta) \frac{\sum_{i=1}^{N_v}d_i^2}{\sum_{i=1}^{N_v}d_i}=-\alpha,
\end{align}
and 
\begin{align*}
\frac{\mathbb E X}{\mathbb E Y}=\bar d +1 -\alpha (2-\alpha-\beta).
\end{align*}
By Theorem 2, for any $\eta>0$, we have  
\begin{align*}
\text{Bias}[\tilde \kappa]=\mathcal O\Big(N_v^{c-\frac{c+1}{2+\eta} }\Big) \text{ as } N_v\rightarrow\infty.
\end{align*}

\subsection{Proof of Corollary 4}
Note that the asymptotic notations for $d_L$ and $\sum_{i=1}^{N_v}d_i^2/\sum_{i=1}^{N_v}d_i$ are same as equation (\ref{eq23}) and equation (\ref{eq24}).

By edge unbiasedness, we have 
\begin{align*}
\alpha=\frac{\beta  \bar d}{N_v-1-\bar d}=\Theta ( N_v^{c-1}).
\end{align*}

(i) $0<\zeta\leq 2$

Note that 
\begin{align*}
\alpha(N_v-1) +\beta -(\alpha+\beta) \frac{\sum_{i=1}^{N_v}d_i^2}{\sum_{i=1}^{N_v}d_i}
= -  \beta \kappa+\alpha(N_v-\kappa-1) +\beta,  
\end{align*}
and 
\begin{align*}
\frac{\mathbb E X}{\mathbb E Y}=\Theta\Big( \frac{\sum_{i=1}^{N_v}d_i^2}{\sum_{i=1}^{N_v}d_i} \Big).
\end{align*}
By Theorem 2, for any $\eta>0$, we have  
\begin{align*}
\text{Bias}[\tilde \kappa]&=-\beta (2-\alpha-\beta)\kappa+ (2-\alpha-\beta) \big[\alpha(N_v-\kappa-1) +\beta   \big]+\mathcal O\Big(\dfrac{1}{(\mathbb E Y)^{1/(2+\eta)}}\dfrac{\mathbb E X}{\mathbb E Y} \Big) \\
&=-\beta (2-\alpha-\beta)\kappa+\mathcal O\Big(\max\Big\{N_v^c,\dfrac{\kappa}{N_v^{(c+1)/(2+\eta)}}\Big \}   \Big) 
\end{align*}
$\text{ as } N_v\rightarrow\infty.$

(ii) $\zeta> 2$

Note that 
\begin{align*}
\alpha(N_v-1) +\beta -(\alpha+\beta) \frac{\sum_{i=1}^{N_v}d_i^2}{\sum_{i=1}^{N_v}d_i}\sim-\dfrac{\beta \bar d}{\zeta(\zeta-2)} -\dfrac{\beta (\bar d)^2}{\zeta(\zeta-2)(N_v-1-\bar d)} +\beta,
\end{align*}
and 
\begin{align*}
\frac{\mathbb E X}{\mathbb E Y}=\mathcal O( \bar d).
\end{align*}
By Theorem 2, for any $\eta>0$, we have  
\begin{align*}
\text{Bias}[\tilde \kappa]&=-  (2-\alpha-\beta)\dfrac{\beta \bar d}{\zeta(\zeta-2)}  +\mathcal O (\max\{N_v^{2c-1},1\})  \\
&=-\beta (2-\alpha-\beta)\dfrac{\kappa}{ (\zeta-1)^2}  +\mathcal O (\max\{N_v^{2c-1},1\}) 
\end{align*}
$\text{ as } N_v\rightarrow\infty.$	

\section{Proofs of theorems for variances of the observed branching number}

To show Theorem \ref{th3} and Theorem \ref{th4}, we first introduce a useful lemma.

\begin{lemma}\label{l1}
	Under assumptions in Theorem \ref{th3}, for any $\eta>0$, we have
	\begin{align*}
	\mathbb E\Big[\Big(\tilde\kappa-\frac{ \mathbb EX}{\mathbb EY}\Big)^2\Big]=&\ \mathbb E\Big[\frac{(X\mathbb EY-Y\mathbb E X)^2}{(\mathbb E Y)^4} \Big]\\
	&+ \mathcal O\Bigg( \max\Big\{  (\mathbb EY)^{-1/(2+\eta)}\cdot\mathbb E\Big[\frac{(X\mathbb EY-Y\mathbb E X)^2}{(\mathbb E Y)^4} \Big],\mathbb P(Y=0)\cdot\Big[\frac{ \mathbb EX}{\mathbb EY}  \Big]^2 \Big\}\Bigg )
	\end{align*}
	as $N_v\rightarrow\infty$.
\end{lemma}
\begin{proof}
	Note that
	\begin{align*}
	&\ \mathbb E\Big[\Big(\tilde\kappa-\frac{ \mathbb EX}{\mathbb EY}\Big)^2\Big]-\mathbb E\Big[\frac{(X\mathbb EY-Y\mathbb E X)^2}{(\mathbb E Y)^4} \Big]\\
	=&\  \mathbb E\Big[\frac{(X\mathbb EY-Y\mathbb E X)^2\big[ (\mathbb EY)^2-Y^2 \big]}{Y^2(\mathbb E Y)^4}\cdot I_{ \{ Y>0 \}}\Big]+\Big(\frac{ \mathbb EX}{\mathbb EY}\Big)^2\cdot\text{Pr}(Y=0).
	\end{align*}
	By triangle inequality and Jensen's inequality, we obtain
	\begin{align*}\label{eq10.5}
	&\ \Bigg| \mathbb E\Big[\Big(\frac{X}{Y}\cdot I_{ \{ Y>0 \}}-\frac{ \mathbb EX}{\mathbb EY}\Big)^2\Big]-\mathbb E\Big[\frac{(X\mathbb EY-Y\mathbb E X)^2}{(\mathbb E Y)^4} \Big]\Bigg|\\
	\leq&\ \frac{1}{(\mathbb E Y)^4}  \mathbb E\Big[\frac{(X\mathbb EY-Y\mathbb E X)^2\big|(\mathbb EY)^2-Y^2 \big|}{Y^2}\cdot I_{ \{ Y>0 \}}\Big] + \Big(\frac{ \mathbb EX}{\mathbb EY}\Big)^2\cdot\text{Pr}(Y=0). \numberthis
	\end{align*}
	
	Next, we find an upper bound of 
	\begin{align}\label{eq10.6}
	\mathbb E\Big[\frac{(X\mathbb EY-Y\mathbb E X)^2\big| (\mathbb EY)^2-Y^2 \big|}{Y^2}\cdot I_{ \{ Y>0 \}}\Big].
	\end{align}
	By additivity of expectation, for any $\delta\in(0,1)$, (\ref{eq10.6}) equals
	\begin{align}\label{eq10.7}
	&\ \mathbb E\Big[\frac{(X\mathbb EY-Y\mathbb E X)^2\big| (\mathbb EY)^2-Y^2 \big|}{Y^2}\cdot I_{ \{ Y>0 \}}\cdot I_{ \{ |Y-\mathbb EY|\geq\delta\mathbb EY  \}}\Big]\\
	&+ \mathbb E\Big[\frac{(X\mathbb EY-Y\mathbb E X)^2\big| (\mathbb EY)^2-Y^2 \big|}{Y^2}\cdot I_{ \{ Y>0 \}}\cdot I_{ \{ |Y-\mathbb EY|<\delta\mathbb EY  \}}\Big].
	\end{align}
	For the first term in (\ref{eq10.7}), by definitions of $X$ and $Y$, we obtain $\Big| \frac{ X\mathbb EY-Y\mathbb E X  }{Y}\Big|\cdot I_{ \{ Y>0 \}}<N_v^3 $ and $| (\mathbb EY)^2-Y^2  | < N_v^4$. Thus, we have
	\begin{align*}
	\mathbb E\Big[\frac{(X\mathbb EY-Y\mathbb E X)^2\big| (\mathbb EY)^2-Y^2 \big|}{Y^2}\cdot I_{ \{ Y>0 \}}\cdot I_{ \{ |Y-\mathbb EY|\geq\delta\mathbb EY  \}}\Big] 
	< \ N_v^{10} \cdot \text{Pr}(|Y-\mathbb EY|\geq\delta\mathbb EY).  
	\end{align*}
	Then, by Chernoff Bound, we obtain
	\begin{align*}
	\mathbb E\Big[\frac{(X\mathbb EY-Y\mathbb E X)^2\big| (\mathbb EY)^2-Y^2 \big|}{Y^2}\cdot I_{ \{ Y>0 \}}\cdot I_{ \{ |Y-\mathbb EY|\geq\delta\mathbb EY  \}}\Big] 
	< 2N_v^{10} \cdot \exp\Big(-\frac{\delta^2\cdot \mathbb EY}{6}\Big).
	\end{align*}
	For the second term in (\ref{eq10.7}), when $|\mathbb EY-Y|< \delta  \mathbb EY$, $Y>(1-\delta) \mathbb EY$ and $Y<(1+\delta) \mathbb EY$. And notice $| (\mathbb EY)^2-Y^2|=(\mathbb EY+Y)|\mathbb EY-Y|$, we have
	\begin{align*}
	&\ \mathbb E\Big[\frac{(X\mathbb EY-Y\mathbb E X)^2\big| (\mathbb EY)^2-Y^2 \big|}{Y^2}\cdot I_{ \{ Y>0 \}}\cdot I_{ \{ |Y-\mathbb EY|<\delta\mathbb EY  \}}\Big] \\
	<  &\ \frac{\delta (2+\delta)}{(1-\delta)^2  }\cdot \mathbb E\Big[ (X\mathbb EY-Y\mathbb E X)^2 \Big].
	\end{align*}
	By (\ref{eq10.5}), we show
	\begin{align*}\label{eq10.9}
	&\ \Bigg| \mathbb E\Big[\Big(\tilde\kappa-\frac{ \mathbb EX}{\mathbb EY}\Big)^2\Big]-\mathbb E\Big[\frac{(X\mathbb EY-Y\mathbb E X)^2}{(\mathbb E Y)^4} \Big]\Bigg|\\
	\leq&\ \frac{2N_v^{10}}{(\mathbb E Y)^4} \cdot \exp\Big(-\frac{\delta^2\cdot \mathbb EY}{6}\Big)+\frac{\delta (2+\delta)}{(1-\delta)^2  }\cdot \mathbb E\Big[\frac{(X\mathbb EY-Y\mathbb E X)^2}{(\mathbb E Y)^4} \Big]+ \Big(\frac{ \mathbb EX}{\mathbb EY}\Big)^2\cdot\text{Pr}(Y=0).\numberthis
	\end{align*}
	
	$L_1(N_v)$ and $L_2(N_v)$ denote the first two terms on the right sides in (\ref{eq10.9}). We choose $\delta=( \mathbb EY)^{-1/(2+\eta)}$, $\eta>0$, such that
	\begin{align*} 
	L_1(N_v) =o\Big( \mathbb E\Big[\frac{(X\mathbb EY-Y\mathbb E X)^2}{(\mathbb E Y)^4} \Big]\Big) \text{ and }L_2(N_v)=o\Big(  \mathbb E\Big[\frac{(X\mathbb EY-Y\mathbb E X)^2}{(\mathbb E Y)^4} \Big]\Big) 
	\end{align*}
	as $N_v\rightarrow \infty$. Under the assumption $\mathbb EY=\Omega(N_v)$ as $N_v\rightarrow\infty$, $1-\delta=\mathcal O(1)$. By L'Hopital's rule, we have 
	\begin{align*}
	L_1(N_v)=o(L_2(N_v))\text{ as }N_v\rightarrow\infty.
	\end{align*}
	These complete the proof.	
\end{proof}

\subsection{Proof of Theorem 3}

By the definition of variance, we have
\begin{align*}
\text{Var}[ \tilde\kappa]=\text{Var}\Big[ \tilde\kappa-\frac{ \mathbb EX}{\mathbb EY}\Big]= \mathbb E\Big[\Big( \tilde\kappa-\frac{ \mathbb EX}{\mathbb EY}\Big)^2\Big]- \Big[ \mathbb E (   \tilde\kappa)-\frac{\mathbb E X}{\mathbb EY}  \Big] ^2.
\end{align*}

(i) By Lemma \ref{l1}, for any $\eta>0$, we obtain	
\begin{align*}
\text{Var}[ \tilde\kappa]&=\mathcal O\Big(\mathbb E\Big[\Big( \tilde\kappa-\frac{ \mathbb EX}{\mathbb EY}\Big)^2\Big]\Big)  \\
&= \mathcal O\Bigg( \max\Big\{   \mathbb E\Big[\frac{(X\mathbb EY-Y\mathbb E X)^2}{(\mathbb E Y)^4} \Big],\mathbb P(Y=0)\cdot\Big[\frac{ \mathbb EX}{\mathbb EY}  \Big]^2 \Big\}\Bigg ).
\end{align*}

(ii) By triangle inequality, we have 

\begin{align*}
&\ \Bigg|\text{Var}[ \tilde\kappa]-\mathbb E\Big[\frac{(X\mathbb EY-Y\mathbb E X)^2}{(\mathbb E Y)^4} \Big]\Bigg| \\
\leq&\  \Bigg| \mathbb E\Big[\Big(\tilde\kappa-\frac{ \mathbb EX}{\mathbb EY}\Big)^2\Big]-\mathbb E\Big[\frac{(X\mathbb EY-Y\mathbb E X)^2}{(\mathbb E Y)^4} \Big] \Bigg|+ \Big[  \mathbb E  (  \tilde\kappa ) -\frac{\mathbb E X}{\mathbb EY}\Big] ^2. 
\end{align*}
Apply Lemma \ref{l1} and Theorem \ref{th1} and the rest follows.

\subsection{Proof of Theorem 4}

Under assumptions in Theorem \ref{th4}, we have 
\begin{align*}
\mathbb P(Y=0)\Big[\frac{ \mathbb EX}{\mathbb EY}  \Big]^2 =o\Big(  \mathbb E\Big[\frac{(X\mathbb EY-Y\mathbb E X)^2}{(\mathbb E Y)^4} \Big]\Big),
\end{align*}
and there exist $\eta_0,\lambda_0>0$, such that
\begin{align*}
\mathbb P(Y=0) \Big[\frac{ \mathbb EX}{\mathbb EY}  \Big]^2 =o \Bigg( \max\Big\{ \frac{1}{(\mathbb EY)^{-1/(2+\eta_0)} }  \mathbb E\Big[\frac{(X\mathbb EY-Y\mathbb E X)^2}{(\mathbb E Y)^4} \Big] ,\frac{1}{(\mathbb EY)^{-2/(2+\lambda_0)} }  \Big[\frac{ \mathbb EX}{\mathbb EY}  \Big]^2  \Big\}\Bigg ).
\end{align*}
Apply Theorem \ref{th3} and the rest follows.

\section{Proofs of corollaries for variances of the observed branching number}

To prove corollaries for variances of the observed branching number, we first compute $\displaystyle\mathbb E\Big[\frac{(X\mathbb EY-Y\mathbb E X)^2}{(\mathbb E Y)^4} \Big]$. Note that 
	\begin{align*}
\mathbb E\Big[\frac{(X\mathbb EY-Y\mathbb E X)^2}{(\mathbb E Y)^4} \Big] =\Big[\frac{ \mathbb EX}{\mathbb EY}  \Big]^2 \cdot \Big[ \frac{\text{Var}X}{ ( \mathbb EX)^2}-2\frac{\text{Cov}(X,Y)}{\mathbb EX\mathbb EY}+ \frac{\text{Var}Y}{ ( \mathbb EY)^2} \Big].
\end{align*}
Under Assumption 1, 2 and 4, we have
\begin{align*}
\mathbb EY=&\ \sum_{i=1}^{N_v} d_i,\\
\mathbb EX=&\  (1-\alpha-\beta)^2\sum_{i=1}^{N_v} d_i^2 +(2-\alpha-\beta)\big[\alpha(N_v-1)+\beta\big] \sum_{i=1}^{N_v} d_i,\\
\text{Var}Y=&\ 2\beta(2-\alpha-\beta)\sum_{i=1}^{N_v} d_i,\\
\text{Cov} (X,Y)=&\ 4 (\beta-\alpha) (1-\alpha-\beta) ^2 \sum_{i=1}^{N_v} d_i^2+ 2  \Big\{ \beta\big[  (1-\alpha )(1-2\alpha)- (1-\beta )(1-2\beta)   \big] \\
&+2\alpha (N_v-1) \big[\beta(1-\beta)+(1-\alpha)^2\big] \Big \} \sum_{i=1}^{N_v} d_i,\\
\text{Var}X=&\ 4 ( \beta-\alpha ) (1 - \alpha - \beta)^3 \sum_{i=1}^{N_v}d_i^3 \\
& +2 (1 - \alpha - \beta)^2  \Big[19 \alpha^2 + 9 \beta^2 - 6 \alpha (1+ 3 \beta) -4\beta+ 
2 \alpha (1 + 4\beta- 5 \alpha ) N_v\Big] \sum_{i=1}^{N_v} d_i^2\\
&+ \Bigg\{(1  -\alpha - \beta)  \Big[-38 \alpha^3 +2 \alpha^2 (17 + 11 \beta) + 
\beta (1 - 6 \beta + 6 \beta^2) - \alpha (5 + 14 \beta^2)\\
& +4 \alpha (1 + 11 \alpha^2 + 2 \beta^2 - \alpha (9 + 5 \beta)) N_v + 
4 \alpha^2 (2 - 3 \alpha + \beta) N_v^2 \Big] \\
&+  (1 - \alpha) (\alpha+\beta) \Big[ 1 - 12 \alpha + 20 \alpha^2 + 6 (1 - 3 \alpha) \alpha N_v + 4 \alpha^2 N_v^2\Big] \\
& +  2  (1-\alpha-\beta) \Big[3 \alpha^2 -  \beta (1 -\beta)- \alpha (1 + 2 \beta) + 
\alpha (1 - 2 \alpha + \beta) N_v\Big] \\
& + 4(1 - \alpha - \beta)\Big[ -8 \alpha^3 +3 \alpha (1 - \beta) \beta + (1 - \beta)^2 \beta + 
4 \alpha^2 (1 + \beta) \Big] \\
&-8 \alpha \Big[ 5 \alpha^3 + (1 - \beta)^2 \beta - \alpha^2 (8 - 3 \beta) +
\alpha (3 - 2 \beta - \beta^2)\Big] N_v \\
&+ 4 \alpha^2 \Big[2 + 3 \alpha^2 - \beta - \beta^2 - \alpha (5 - 2 \beta)\Big]N_v^2\\
&+2\alpha(1-\alpha)(\alpha+\beta)(N_v-2) \Big[ 1+2\alpha(N_v-2)  \Big] \\
& +\beta\Big[  \alpha(\alpha-3)- \beta(\beta-3)   \Big] 
+2\alpha (N_v-1) \Big[\beta(1-\beta)+(1-\alpha)^2\Big] \Bigg \} \sum_{i=1}^{N_v} d_i\\
&+ 4  (1-\alpha-\beta)^2\Big[ \alpha (1-\alpha) \sum_{i=1}^{N_v} \sum_{j\neq i} d_id_j I_{ \{A_{ij}=0\} } \\
&+\beta(1-\beta)\sum_{i=1}^{N_v} \sum_{j\neq i} d_id_j I_{ \{A_{ij}=1\} }\Big].
\end{align*}

\subsection{Proof of Corollary 5}

By edge unbiasedness, we have 
\begin{align}
\alpha=\frac{\beta  \bar d}{N_v-1-\bar d}=\Theta\Big(\frac{\log N_v}{N_v}\Big).
\end{align}
By homogeneity, we obtain
\begin{align*}
\sum_{i=1}^{N_v}d_i^2&=\Theta(  N_v \log^2 (N_v) ),\\
\sum_{i=1}^{N_v}d_i^3&=\Theta( N_v \log^3 (N_v)  ).
\end{align*}
Besides, by Young's inequality, we have
\begin{align*}
\sum_{i=1}^{N_v} \sum_{j\neq i} d_id_j I_{\{A_{ij}=1 \}} &\leq \Big(\sum_{i=1}^{N_v} \sum_{j\neq i} d_i^2d_j^2\Big)^{1/2}\cdot \Big(\sum_{i=1}^{N_v} \sum_{j\neq i} I_{\{A_{ij}=1\}}\Big)^{1/2} =\mathcal{O} (N_v^{3/2}\log^{5/2} N_v),\\
\sum_{i=1}^{N_v} \sum_{j\neq i} d_id_j I_{\{A_{ij}=0 \}} &\leq \Big(\sum_{i=1}^{N_v} \sum_{j\neq i} d_i^2d_j^2\Big)^{1/2}\cdot \Big(\sum_{i=1}^{N_v} \sum_{j\neq i} I_{\{A_{ij}=0\}}\Big)^{1/2} =\mathcal{O} (N_v^2\log^2 N_v).
\end{align*}
Thus, we obtain
\begin{align*}
\frac{ \mathbb EX}{\mathbb EY}&= \Theta(\log N_v),\\
\frac{\text{Var}X}{ ( \mathbb EX)^2}&=\mathcal O\Big(\frac{1}{N_v^{1/2} \log^{3/2} N_v}\Big),\\
\frac{\text{Cov}(X,Y)}{\mathbb EX\mathbb EY}&=\Theta\Big(\frac{1}{N_v\log N_v}\Big),\\
\frac{\text{Var}Y}{ ( \mathbb EY)^2}&=\Theta\Big(\frac{1}{N_v\log N_v}\Big).
\end{align*}
Then, we have
\begin{align*}
\mathbb E\Big[\frac{(X\mathbb EY-Y\mathbb E X)^2}{(\mathbb E Y)^4} \Big] =\mathcal{O}\Bigg( \Big( \frac{\log N_v}{N_v}\Big)^{1/2}\Bigg).
\end{align*}
By Theorem 4,   
\begin{align*}
\text{Var}[\tilde \kappa]= \mathcal{O}\Bigg( \Big( \frac{\log N_v}{N_v}\Big)^{1/2}\Bigg)\text{ as } N_v\rightarrow\infty.
\end{align*}

\subsection{Proof of Corollary 6}

By edge unbiasedness, we have 
\begin{align}
\alpha=\frac{\beta  \bar d}{N_v-1-\bar d}=\Theta\Big(\frac{\log N_v}{N_v}\Big).
\end{align}
Next, we compute $\sum_{i=1}^{N_v} d_i^3$ for different values of $\zeta$.  
\begin{align*}  \label{eq18}
\int_{d_L}^{N_v-1}x^3\cdot \frac{\zeta d_L^\zeta }{1- \Big(\dfrac{d_L}{N_v-1}\Big )^\zeta } x^{-(\zeta+1)}dx
&=\begin{cases}
\zeta d_L^\zeta \cdot  \dfrac{\log\Big(\dfrac{N_v-1}{d_L}\Big ) } {1- \Big(\dfrac{d_L}{N_v-1}\Big )^\zeta } ,   \text{ if }  \zeta=3\\
\dfrac{\zeta d_L^3}{\zeta-3}\cdot  \dfrac{1- \Big(\dfrac{d_L}{N_v-1}\Big )^{\zeta-3} } {1- \Big(\dfrac{d_L}{N_v-1}\Big )^\zeta } ,   \text{ otherwise. }   \\
\end{cases}\numberthis
\end{align*}
Thus,  we obtain
\begin{align*}
\sum_{i=1}^{N_v} d_i^3=\begin{cases}
\Theta(N_v^3\log N_v), &\text{ if } 0<\zeta<1\\
\Theta(N_v^3), &\text{ if } \zeta=1 \\
\Theta(N_v^{4-\zeta}\cdot \log  ^{\zeta }N_v ), &\text{ if } 1<\zeta<3 \\
\Theta(N_v \log^4 N_v), &\text{ if } \zeta=3\\
\Theta( N_v \log^3 N_v), &\text{ if } \zeta>3.\\
\end{cases}
\end{align*}

Equation (\ref{eq10}) leads to
\begin{align*}
\sum_{i=1}^{N_v} d_i^2=\begin{cases}
\Theta(N_v^2\log N_v), &\text{ if } 0<\zeta<1 \\
\Theta(N_v^2), &\text{ if } \zeta=1 \\
\Theta(N_v^{3-\zeta}\cdot \log  ^{\zeta}N_v ), &\text{ if } 1<\zeta<2 \\
\Theta( N_v\log^3 N_v), &\text{ if } \zeta=2 \\
\Theta(N_v \log^2 N_v), &\text{ if } \zeta>2. \\
\end{cases}
\end{align*}
In addition, by Young's inequality, we have 
\begin{align*} \label{eq19}
\sum_{i=1}^{N_v} \sum_{j\neq i} d_id_j I_{\{A_{ij}=1 \}} &\leq \Big(\sum_{i=1}^{N_v} \sum_{j\neq i} d_i^2d_j^2\Big)^{1/2}\cdot \Big(\sum_{i=1}^{N_v} \sum_{j\neq i} I_{\{A_{ij}=1\}}\Big)^{1/2},\\
\sum_{i=1}^{N_v} \sum_{j\neq i} d_id_j I_{\{A_{ij}=0 \}} &\leq \Big(\sum_{i=1}^{N_v} \sum_{j\neq i} d_i^2d_j^2\Big)^{1/2}\cdot \Big(\sum_{i=1}^{N_v} \sum_{j\neq i} I_{\{A_{ij}=0\}}\Big)^{1/2}. \numberthis
\end{align*}
Thus, we obtain
\begin{align*}
\sum_{i=1}^{N_v} \sum_{j\neq i} d_id_j I_{\{A_{ij}=1 \}} =\begin{cases}
\mathcal O(N_v^{5/2}\log^{3/2} N_v), &\text{ if } 0<\zeta<1 \\
\mathcal O(N_v^{5/2}\log^{1/2} N_v), &\text{ if } \zeta=1 \\
\mathcal O(N_v^{7/2-\zeta}\cdot \log  ^{\zeta+1/2}N_v ), &\text{ if } 1<\zeta<2 \\
\mathcal O( N_v^{3/2}\log^{7/2}N_v), &\text{ if } \zeta=2 \\
\mathcal O(N_v^{3/2} \log^{5/2} N_v), &\text{ if } \zeta>2,
\end{cases}
\end{align*}
and
\begin{align*}
\sum_{i=1}^{N_v} \sum_{j\neq i} d_id_j I_{\{A_{ij}=0 \}} =\begin{cases}
\mathcal O(N_v^3\log N_v), &\text{ if } 0<\zeta<1\\
\mathcal O(N_v^3), &\text{ if } \zeta=1 \\
\mathcal O(N_v^{4-\zeta}\cdot \log  ^{\zeta }N_v ), &\text{ if } 1<\zeta<2 \\
\mathcal O( N_v^2\log^3 N_v), &\text{ if } \zeta=2 \\
\mathcal O(N_v^2 \log^2 N_v), &\text{ if } \zeta>2.
\end{cases}
\end{align*}

(i) $0<\zeta<1$

Note that 
\begin{align*}
\frac{ \mathbb EX}{\mathbb EY}&= \Theta\Big(N_v\Big),\\
\frac{\text{Var}X}{ ( \mathbb EX)^2}&=\Theta\Big(\frac{1}{N_v\log N_v}\Big),\\
\frac{\text{Cov}(X,Y)}{\mathbb EX\mathbb EY}&=\Theta\Big(\frac{1}{N_v\log N_v}\Big),\\
\frac{\text{Var}Y}{ ( \mathbb EY)^2}&=\Theta\Big(\frac{1}{N_v\log N_v}\Big).
\end{align*}
Then, we have
\begin{align*}
\mathbb E\Big[\frac{(X\mathbb EY-Y\mathbb E X)^2}{(\mathbb E Y)^4} \Big] =\mathcal{O}\Big(\frac{ N_v}{\log N_v}\Big).
\end{align*}
By Theorem 4,   
\begin{align*}
\text{Var}[\tilde \kappa]= \mathcal{O}\Big(\frac{ N_v}{\log N_v}\Big)\text{ as } N_v\rightarrow\infty.
\end{align*}

(ii) $ \zeta=1$

Note that 
\begin{align*}
\frac{ \mathbb EX}{\mathbb EY}&= \Theta\Big(\frac{N_v}{\log N_v}\Big),\\
\frac{\text{Var}X}{ ( \mathbb EX)^2}&=\Theta\Big(\frac{1}{N_v}\Big),\\
\frac{\text{Cov}(X,Y)}{\mathbb EX\mathbb EY}&=\Theta\Big(\frac{1}{N_v\log N_v}\Big),\\
\frac{\text{Var}Y}{ ( \mathbb EY)^2}&=\Theta\Big(\frac{1}{N_v\log N_v}\Big).
\end{align*}
Then, we have
\begin{align*}
\mathbb E\Big[\frac{(X\mathbb EY-Y\mathbb E X)^2}{(\mathbb E Y)^4} \Big] =\mathcal{O}\Big(\frac{ N_v}{\log^2 N_v}\Big).
\end{align*}
By Theorem 4,   
\begin{align*}
\text{Var}[\tilde \kappa]= \mathcal{O}\Big(\frac{ N_v}{\log^2 N_v}\Big)\text{ as } N_v\rightarrow\infty.
\end{align*}

(iii) $1<\zeta<2$

Note that 
\begin{align*}
\frac{ \mathbb EX}{\mathbb EY}&= \Theta\Big(N_v^{2-\zeta }\cdot  \log ^{\zeta -1} N_v\Big),\\
\frac{\text{Var}X}{ ( \mathbb EX)^2}&=\Theta\Big( \frac{1}{N_v^{2-\zeta }\cdot  \log ^{\zeta  } N_v} \Big),\\
\frac{\text{Cov}(X,Y)}{\mathbb EX\mathbb EY}&=\Theta\Big(\frac{1}{N_v\log N_v}\Big),\\
\frac{\text{Var}Y}{ ( \mathbb EY)^2}&=\Theta\Big(\frac{1}{N_v\log N_v}\Big).
\end{align*}
Then, we have
\begin{align*}
\mathbb E\Big[\frac{(X\mathbb EY-Y\mathbb E X)^2}{(\mathbb E Y)^4} \Big] =\mathcal{O}\Big(  \Big(\frac{ N_v}{\log N_v} \Big)^{2-\zeta } \Big).
\end{align*}
By Theorem 4,   
\begin{align*}
\text{Var}[\tilde \kappa]=\mathcal{O}\Big(\Big(\frac{ N_v}{\log N_v} \Big)^{2-\zeta }\Big)\text{ as } N_v\rightarrow\infty.
\end{align*}

(iv) $\zeta=2$

Note that 
\begin{align*}
\frac{ \mathbb EX}{\mathbb EY}&= \Theta\Big(\log ^{2} N_v\Big),\\
\frac{\text{Var}X}{ ( \mathbb EX)^2}&=\Theta\Big( \frac{1}{\log ^{4} N_v} \Big),\\
\frac{\text{Cov}(X,Y)}{\mathbb EX\mathbb EY}&=\Theta\Big(\frac{1}{N_v\log N_v}\Big),\\
\frac{\text{Var}Y}{ ( \mathbb EY)^2}&=\Theta\Big(\frac{1}{N_v\log N_v}\Big).
\end{align*}
Then, we have
\begin{align*}
\mathbb E\Big[\frac{(X\mathbb EY-Y\mathbb E X)^2}{(\mathbb E Y)^4} \Big] =\mathcal{O}\Big(  1 \Big).
\end{align*}
By Theorem 4,   
\begin{align*}
\text{Var}[\tilde \kappa]= \mathcal{O}\Big(1\Big)\text{ as } N_v\rightarrow\infty.
\end{align*}

(v) $2<\zeta<5/2$

Note that 
\begin{align*}
\frac{ \mathbb EX}{\mathbb EY}&= \Theta\Big(  \log N_v\Big),\\
\frac{\text{Var}X}{ ( \mathbb EX)^2}&=\Theta\Big( \frac{1}{ N_v^{\zeta-2 } \log ^{4-\zeta } N_v} \Big),\\
\frac{\text{Cov}(X,Y)}{\mathbb EX\mathbb EY}&=\Theta\Big(\frac{1}{N_v\log N_v}\Big),\\
\frac{\text{Var}Y}{ ( \mathbb EY)^2}&=\Theta\Big(\frac{1}{N_v\log N_v}\Big).
\end{align*}
Then, we have
\begin{align*}
\mathbb E\Big[\frac{(X\mathbb EY-Y\mathbb E X)^2}{(\mathbb E Y)^4} \Big] =\mathcal{O}\Big(\Big(\frac{\log N_v}{ N_v} \Big)^{\zeta-2 }\Big).
\end{align*}
By Theorem 4,   
\begin{align*}
\text{Var}[\tilde \kappa]= \mathcal{O}\Big(\Big(\frac{\log N_v}{ N_v} \Big)^{\zeta-2 }\Big)\text{ as } N_v\rightarrow\infty.
\end{align*}

(vi) $\zeta\geq 5/2$

Note that 
\begin{align*}
\frac{ \mathbb EX}{\mathbb EY}&= \Theta\Big(  \log N_v\Big),\\
\frac{\text{Var}X}{ ( \mathbb EX)^2}&=\mathcal O\Big( \frac{1}{ N_v^{1/2 } \log ^{3/2 } N_v} \Big),\\
\frac{\text{Cov}(X,Y)}{\mathbb EX\mathbb EY}&=\Theta\Big(\frac{1}{N_v\log N_v}\Big),\\
\frac{\text{Var}Y}{ ( \mathbb EY)^2}&=\Theta\Big(\frac{1}{N_v\log N_v}\Big).
\end{align*}
Then, we have
\begin{align*}
\mathbb E\Big[\frac{(X\mathbb EY-Y\mathbb E X)^2}{(\mathbb E Y)^4} \Big] =\mathcal{O}\Big(\Big(\frac{\log N_v}{ N_v} \Big)^{1/2 }\Big).
\end{align*}
By Theorem 4,   
\begin{align*}
\text{Var}[\tilde \kappa]= \mathcal{O}\Big(\Big(\frac{\log N_v}{ N_v} \Big)^{1/2 }\Big)\text{ as } N_v\rightarrow\infty.
\end{align*}

\subsection{Proof of Corollary 7}

By edge unbiasedness, we have 
\begin{align}
\alpha=\frac{\beta  \bar d}{N_v-1-\bar d}=\Theta(N_v^{c-1}).
\end{align}
By homogeneity, we obtain
\begin{align*}
\sum_{i=1}^{N_v}d_i^2&=\Theta(N_v^{2c+1}),\\
\sum_{i=1}^{N_v}d_i^3&=\Theta( N_v ^{3c+1}).
\end{align*}
Besides, by Young's inequality, we have
\begin{align*}
\sum_{i=1}^{N_v} \sum_{j\neq i} d_id_j I_{\{A_{ij}=1 \}} &\leq \Big(\sum_{i=1}^{N_v} \sum_{j\neq i} d_i^2d_j^2\Big)^{1/2}\cdot \Big(\sum_{i=1}^{N_v} \sum_{j\neq i} I_{\{A_{ij}=1\}}\Big)^{1/2} =\mathcal{O} (N_v^{(5c+3)/2}),\\
\sum_{i=1}^{N_v} \sum_{j\neq i} d_id_j I_{\{A_{ij}=0 \}} &\leq \Big(\sum_{i=1}^{N_v} \sum_{j\neq i} d_i^2d_j^2\Big)^{1/2}\cdot \Big(\sum_{i=1}^{N_v} \sum_{j\neq i} I_{\{A_{ij}=0\}}\Big)^{1/2} =\mathcal{O} (N_v^{2c+2}).
\end{align*}
Thus, we obtain
\begin{align*}
\frac{ \mathbb EX}{\mathbb EY}&= \Theta( N_v^c),\\
\frac{\text{Var}X}{ ( \mathbb EX)^2}&=\mathcal O\Big(\frac{1}{N_v^{(3c+1)/2} }\Big),\\
\frac{\text{Cov}(X,Y)}{\mathbb EX\mathbb EY}&=\Theta\Big(\frac{1}{N_v^{c+1}}\Big),\\
\frac{\text{Var}Y}{ ( \mathbb EY)^2}&=\Theta\Big(\frac{1}{N_v^{c+1}}\Big).
\end{align*}
Then, we have
\begin{align*}
\mathbb E\Big[\frac{(X\mathbb EY-Y\mathbb E X)^2}{(\mathbb E Y)^4} \Big] =\mathcal{O}(N_v^{(c-1)/2}).
\end{align*}
By Theorem 4,   
\begin{align*}
\text{Var}[\tilde \kappa]=\mathcal{O}(N_v^{(c-1)/2})\text{ as } N_v\rightarrow\infty.
\end{align*}

\subsection{Proof of Corollary 8}

By edge unbiasedness, we have 
\begin{align}
\alpha=\frac{\beta  \bar d}{N_v-1-\bar d}=\Theta(N_v^{c-1}).
\end{align}
By equation (\ref{eq18}), we have  
\begin{align*}
\sum_{i=1}^{N_v} d_i^3=\begin{cases}
\Theta(N_v^{c+3}), &\text{ if } 0<\zeta<1  \\
\Theta(N_v^{c+3}/\log N_v), &\text{ if } \zeta=1 \\
\Theta(N_v^{4-\zeta+c\zeta } ), &\text{ if } 1<\zeta<3 \\
\Theta( N_v^{3c+1}\cdot\log N_v), &\text{ if } \zeta=3\\
\Theta( N_v^{3c+1}), &\text{ if } \zeta>3. \\
\end{cases}
\end{align*}
Equation (\ref{eq10}) leads to
\begin{align*}
\sum_{i=1}^{N_v} d_i^2=\begin{cases}
\Theta(N_v^{c+2}), &\text{ if } 0<\zeta<1  \\
\Theta(N_v^{c+2}/\log N_v), &\text{ if } \zeta=1 \\
\Theta(N_v^{3-\zeta+c\zeta} ), &\text{ if } 1<\zeta<2 \\
\Theta( N_v^{2c+1}\cdot\log N_v), &\text{ if } \zeta=2 \\
\Theta( N_v^{2c+1}), &\text{ if } \zeta>2.\\
\end{cases}
\end{align*}
In addition, by equation (\ref{eq19}), we obtain
\begin{align*}
\sum_{i=1}^{N_v} \sum_{j\neq i} d_id_j I_{\{A_{ij}=1 \}} =\begin{cases}
\mathcal O(N_v^{(3c+5)/2}), &\text{ if } 0<\zeta<1 \\
\mathcal O(N_v^{(3c+5)/2}/\log N_v), &\text{ if } \zeta=1 \\
\mathcal O(N_v^{7/2-\zeta+c(\zeta+1/2)} ), &\text{ if } 1<\zeta<2\\
\mathcal O( N_v^{(5c+3)/2}\cdot\log N_v), &\text{ if } \zeta=2 \\
\mathcal O( N_v^{(5c+3)/2}), &\text{ if } \zeta>2, \\
\end{cases}
\end{align*}
and
\begin{align*}
\sum_{i=1}^{N_v} \sum_{j\neq i} d_id_j I_{\{A_{ij}=0 \}} =\begin{cases}
\mathcal O(N_v^{c+3}), &\text{ if } 0<\zeta<1 \\
\mathcal O(N_v^{c+3}/\log N_v), &\text{ if } \zeta=1 \\
\mathcal O(N_v^{4-\zeta+c\zeta} ), &\text{ if } 1<\zeta<2 \\
\mathcal O( N_v^{2c+2}\cdot\log N_v), &\text{ if } \zeta=2 \\
\mathcal O( N_v^{2c+2}), &\text{ if } \zeta>2. \\
\end{cases}
\end{align*}

(i) $0<\zeta<1$

Note that 
\begin{align*}
\frac{ \mathbb EX}{\mathbb EY}&= \Theta\Big(N_v\Big),\\
\frac{\text{Var}X}{ ( \mathbb EX)^2}&=\Theta\Big(\frac{1}{N_v^{c+1} }\Big),\\
\frac{\text{Cov}(X,Y)}{\mathbb EX\mathbb EY}&=\Theta\Big(\frac{1}{N_v^{c+1} }\Big),\\
\frac{\text{Var}Y}{ ( \mathbb EY)^2}&=\Theta\Big(\frac{1}{N_v^{c+1} }\Big).
\end{align*}
Then, we have
\begin{align*}
\mathbb E\Big[\frac{(X\mathbb EY-Y\mathbb E X)^2}{(\mathbb E Y)^4} \Big] =\mathcal{O}\Big(N_v^{1-c}\Big).
\end{align*}
By Theorem 4,   
\begin{align*}
\text{Var}[\tilde \kappa]=\mathcal{O}\Big(N_v^{1-c}\Big)\text{ as } N_v\rightarrow\infty.
\end{align*}

(ii) $ \zeta=1$

Note that 
\begin{align*}
\frac{ \mathbb EX}{\mathbb EY}&= \Theta\Big(\frac{N_v}{\log N_v}\Big),\\
\frac{\text{Var}X}{ ( \mathbb EX)^2}&=\Theta\Big(\frac{\log N_v}{N_v^{c+1}}\Big),\\
\frac{\text{Cov}(X,Y)}{\mathbb EX\mathbb EY}&=\Theta\Big(\frac{1}{N_v^{c+1}}\Big),\\
\frac{\text{Var}Y}{ ( \mathbb EY)^2}&=\Theta\Big(\frac{1}{N_v^{c+1}}\Big).
\end{align*}
Then, we have
\begin{align*}
\mathbb E\Big[\frac{(X\mathbb EY-Y\mathbb E X)^2}{(\mathbb E Y)^4} \Big] =\mathcal{O}\Big(\frac{ N_v^{1-c}}{\log N_v}\Big).
\end{align*}
By Theorem 4,   
\begin{align*}
\text{Var}[\tilde \kappa]=\mathcal{O}\Big(\frac{ N_v^{1-c}}{\log N_v}\Big)\text{ as } N_v\rightarrow\infty.
\end{align*}

(iii) $1<\zeta<2$

Note that 
\begin{align*}
\frac{ \mathbb EX}{\mathbb EY}&= \Theta\Big(N_v^{2-\zeta+c(\zeta-1)}\Big),\\
\frac{\text{Var}X}{ ( \mathbb EX)^2}&=\Theta\Big( \frac{1}{N_v^{2-\zeta+c\zeta }} \Big),\\
\frac{\text{Cov}(X,Y)}{\mathbb EX\mathbb EY}&=\Theta\Big(\frac{1}{N_v^{c+1}}\Big),\\
\frac{\text{Var}Y}{ ( \mathbb EY)^2}&=\Theta\Big(\frac{1}{N_v^{c+1}}\Big).
\end{align*}
Then, we have
\begin{align*}
\mathbb E\Big[\frac{(X\mathbb EY-Y\mathbb E X)^2}{(\mathbb E Y)^4} \Big] =\mathcal{O}\Big( N_v^{(2-\zeta)(1-c) } \Big).
\end{align*}
By Theorem 4,   
\begin{align*}
\text{Var}[\tilde \kappa]= \mathcal{O}\Big( N_v^{(2-\zeta)(1-c) } \Big)\text{ as } N_v\rightarrow\infty.
\end{align*}

(iv) $\zeta=2$

Note that 
\begin{align*}
\frac{ \mathbb EX}{\mathbb EY}&= \Theta\Big(N_v^{c}\cdot\log N_v\Big),\\
\frac{\text{Var}X}{ ( \mathbb EX)^2}&=\Theta\Big( \frac{1}{N_v^{2c}\cdot\log^2 N_v} \Big),\\
\frac{\text{Cov}(X,Y)}{\mathbb EX\mathbb EY}&=\Theta\Big(\frac{1}{N_v^{c+1}}\Big),\\
\frac{\text{Var}Y}{ ( \mathbb EY)^2}&=\Theta\Big(\frac{1}{N_v^{c+1}}\Big).
\end{align*}
Then, we have
\begin{align*}
\mathbb E\Big[\frac{(X\mathbb EY-Y\mathbb E X)^2}{(\mathbb E Y)^4} \Big] =\mathcal{O}\Big( 1 \Big).
\end{align*}
By Theorem 4,   
\begin{align*}
\text{Var}[\tilde \kappa]= \mathcal{O}\Big( 1\Big)\text{ as } N_v\rightarrow\infty.
\end{align*}

(v) $2<\zeta<5/2$

Note that 
\begin{align*}
\frac{ \mathbb EX}{\mathbb EY}&= \Theta\Big(N_v^{c}\Big),\\
\frac{\text{Var}X}{ ( \mathbb EX)^2}&=\Theta\Big( \frac{1}{N_v^{\zeta-2+c(4-\zeta) }} \Big),\\
\frac{\text{Cov}(X,Y)}{\mathbb EX\mathbb EY}&=\Theta\Big(\frac{1}{N_v^{c+1}}\Big),\\
\frac{\text{Var}Y}{ ( \mathbb EY)^2}&=\Theta\Big(\frac{1}{N_v^{c+1}}\Big).
\end{align*}
Then, we have
\begin{align*}
\mathbb E\Big[\frac{(X\mathbb EY-Y\mathbb E X)^2}{(\mathbb E Y)^4} \Big] =\mathcal{O}\Big( N_v^{(2-\zeta)(1-c) } \Big).
\end{align*}
By Theorem 4,   
\begin{align*}
\text{Var}[\tilde \kappa]= \mathcal{O}\Big( N_v^{(2-\zeta)(1-c) } \Big)\text{ as } N_v\rightarrow\infty.
\end{align*}

(vi) $\zeta\geq 5/2$

Note that 
\begin{align*}
\frac{ \mathbb EX}{\mathbb EY}&= \Theta\Big(N_v^{c}\Big),\\
\frac{\text{Var}X}{ ( \mathbb EX)^2}&=\mathcal O\Big( \frac{1}{N_v^{(3c+1)/2}} \Big),\\
\frac{\text{Cov}(X,Y)}{\mathbb EX\mathbb EY}&=\Theta\Big(\frac{1}{N_v^{c+1}}\Big),\\
\frac{\text{Var}Y}{ ( \mathbb EY)^2}&=\Theta\Big(\frac{1}{N_v^{c+1}}\Big).
\end{align*}
Then, we have
\begin{align*}
\mathbb E\Big[\frac{(X\mathbb EY-Y\mathbb E X)^2}{(\mathbb E Y)^4} \Big] =\mathcal{O}\Big( N_v^{(c-1)/2 } \Big).
\end{align*}
By Theorem 4,   
\begin{align*}
\text{Var}[\tilde \kappa]= \mathcal{O}\Big( N_v^{(c-1)/2 } \Big)\text{ as } N_v\rightarrow\infty.
\end{align*}

\section{Proofs of theorem for the method-of-moments estimator $\hat\kappa$}
\cite{chang2020estimation} provide joint inference of higher-order subgraph densities with unknown error rates. Mimicking their proofs, we can easily obtain the asymptotic joint normal distribution of $\hat C_{\mathcal V_1}$ and $\hat C_{\mathcal V_2}$. Then, by the delta method, we can derive the asymptotic normal distribution of $\hat \kappa$. 
\section{Algorithm for estimation of asymptotic variance of method-of-moments estimator $\hat\kappa$}

To evaluate the asymptotic variance of method-of-moments estimator $\hat\kappa$, we first estimate the asymptotic variance of $(\hat C_{\mathcal V_1}, \hat C_{\mathcal V_2})$ by the method in Section 4 of \cite{chang2020estimation}. Then, we use the delta method to obtain the estimation of the asymptotic variance of $\hat\kappa$. The detail is shown in Algorithm \ref{algo2}.

\begin{algorithm}[!h]     
	\caption{Estimation of asymptotic variance of method-of-moments estimator $\hat \kappa$} 
	\hspace*{0.02in} {\bf Input:} 
	$\tilde {\bm A}=(\tilde A_{i,j})_{N_v\times N_v},\ \varepsilon,\ N_b$, $\hat\alpha$, $\hat\beta$, $\hat k_3$, $\hat C_{\mathcal V_1}$, $\hat C_{\mathcal V_2}$, $\hat\delta$\\
	\hspace*{0.02in} {\bf Output:} 
	$\widehat{\text{Var}}(\hat\kappa)$                
	\begin{algorithmic} 
		\If{$|\hat \alpha-\hat\beta|<\varepsilon$ }
		\State $\xi_2=\hat\alpha,\ \xi_1=1-2\xi_2$; 
		\EndIf
		\If{$\hat\beta-\hat \alpha>\varepsilon$ }
		\State $t_1=\sqrt{1-4\hat \alpha(1-\hat\beta)},\ t_2=\sqrt{1-4\hat\beta (1-\hat\alpha)},\ \xi_2=(1-t_1)/2$;
		\If{$t_1+t_2<0.5$}
		\State $\xi_1=(t_1+t_2)/2$;
		\Else
		\State $\xi_1=(t_1-t_2)/2$;
		\EndIf
		\EndIf
		\If{$\hat \alpha-\hat\beta>\varepsilon$ }
		\State $t_1=\sqrt{1-4\hat \alpha(1-\hat\beta)},\ t_2=\sqrt{1-4\hat\beta (1-\hat\alpha)},\ \xi_2=(1+t_1)/2,\ \xi_1=(t_2-t_1)/2$;
		\EndIf
		\For{$n_b= 1:N_b$}
		\For{$i=1:N_v$}
		\For{$j=i+1:N_v$}
		\State Draw $\eta_{i,j}$ from distribution $\mathbb P(\eta_{i,j}=0)=\xi_1,\ \mathbb P(\eta_{i,j}=1)=\xi_2$ and $\mathbb P(\eta_{i,j}=-1)=1-\xi_1-\xi_2$;
		\State Compute $\tilde A_{i,j}^\dag=\tilde A_{i,j} I(\eta_{i,j}=0)+I(\eta_{i,j}=1)$;
		\State Compute $\mathring{\tilde A}_{i,j}^\dag=\mathring{\tilde A}_{j,i}^\dag=\tilde A_{i,j}^\dag- \tilde A_{i,j} \xi_1-\xi_2$;
		\EndFor
		\EndFor
		\vspace{.2cm}\State $ \hat S_{\mathcal V_1,n_b}^\dag\gets\frac{1}{\hat  k_3}\sqrt{\frac{2}{N_v(N_v-1)}}\ \sum_{i<j}\mathring{\tilde A}_{i,j}^\dag$;
		\vspace{.2cm}\State $  \hat S_{\mathcal V_2,n_b}^\dag\gets\frac{1}{\hat  k_3^2(N_v-2)}\sqrt{\frac{1}{2N_v(N_v-1)}}\ \sum_{i\neq j \neq l}\{ \mathring{\tilde A}_{i,j}^\dag(\tilde A_{j,l} -\hat\alpha)+\mathring{\tilde A}_{j,l}^\dag(\tilde A_{i,j} -\hat\alpha) \}$;
		\EndFor
		\State Compute $\hat{\bm{V} }_{N_v}=\hat{\bm{V} }_{1,N_v}+\hat{\bm{V} }_{2,N_v}+\hat{\bm{V} }_{3,N_v}$, 
		\State	\hspace{1.55cm} where $\hat{\bm{V} }_{1,N_v}=\begin{bmatrix} 
		\text{Var}(\hat S_{\mathcal V_1}^\dag) & \text{Cov}(\hat S_{\mathcal V_1}^\dag,\hat S_{\mathcal V_2}^\dag)  \\
		\text{Cov}(\hat S_{\mathcal V_1}^\dag,\hat S_{\mathcal V_2}^\dag) & \text{Var}(\hat S_{\mathcal V_2}^\dag) 
		\end{bmatrix}$, $\hat{\bm{V} }_{2,N_v}=\hat {\bm\Delta}\hat {\bm G}\hat {\bm\Sigma}\hat {\bm G}^\top\hat {\bm\Delta}^\top$, 
		\State	\hspace{1.55cm} $\hat{\bm{V} }_{3,N_v}=(\hat {\bm H}\hat{\bm G}^\top\hat {\bm\Delta}^\top+\hat {\bm\Delta}\hat {\bm G}\hat {\bm H}^\top)/2$, $\hat {\bm\Delta},\ \hat {\bm G},\ \hat {\bm\Sigma}$, and $\hat{\bm H}$ defined in (\ref{eq10.10});
		\State Compute $\widehat{\text{Var}}(\hat\kappa)=(N_v-2)^2[-\hat C_{\mathcal V_2}/\hat C_{\mathcal V_1}^2,1/\hat C_{\mathcal V_1}] \hat{\bm{V} }_{N_v} [-\hat C_{\mathcal V_2}/\hat C_{\mathcal V_1}^2,1/\hat C_{\mathcal V_1}]^\top$.   
	\end{algorithmic}
	\label{algo2}
\end{algorithm}

\begin{align*}\label{eq10.10}
\hat {\bm\Sigma}= &\ (\hat\sigma_{ij})_{3\times3},\\
\hat {\bm\Delta}=&\ \hat k_3^{-1}\cdot\begin{pmatrix}
\hat C_{\mathcal V_1}-1&  \hat C_{\mathcal V_1}\\
2 \hat C_{\mathcal V_2}-2 \hat C_{\mathcal V_1}& 2 \hat C_{\mathcal V_2}
\end{pmatrix},\\
\hat {\bm G}=&\ \hat k_3^{-2}\cdot\begin{pmatrix}
(1-\hat\delta)^{-1} \{ (1-2\hat\beta)\hat\alpha+\hat\beta^2 \} &(1-\hat\delta)^{-1} (\hat\alpha-2\hat\beta) &(1-\hat\delta)^{-1} \\
-\hat\delta^{-1} \{ (1-2\hat\alpha)\hat\beta+\hat\alpha^2 \} &\hat\delta^{-1}  (\hat\beta-2\hat\alpha+1) &-\hat\delta^{-1} 
\end{pmatrix},\\
\hat{\bm H}=&\ \frac{1}{3}\cdot\begin{pmatrix}
\hat C_{\mathcal V_1}  & \hat k_3^{-1}\\
2\hat C_{\mathcal V_2}   &2\hat k_3^{-1} \hat C_{\mathcal V_1}\\
\end{pmatrix}\\
&\ \cdot\begin{pmatrix}
6\hat k_4 &3(\hat k_4^2-\hat k_1-\hat k_2) & 2\{\hat k_4(-6\hat\alpha\hat\beta+3\hat k_3^2-4\hat k_3)+(1-\hat\alpha)(\hat\beta-2\hat\alpha)\}\\
6\hat k_1 &3\hat k_1(1-2\hat\alpha) & 2\hat k_1(1-\hat\alpha)(1-3\hat\alpha)
\end{pmatrix},\numberthis
\end{align*}
where $\hat k_1=\hat\alpha(1-\hat\alpha)$, $\hat k_2=\hat\beta(1-\hat\beta)$, $\hat k_3=1-\hat\alpha-\hat\beta$, $\hat k_4=\hat\beta- \hat\alpha$, $\hat\sigma_{11}=\hat\delta\hat k_2+(1-\hat\delta)\hat k_1,\ \hat\sigma_{22}=\hat\delta\hat k_2(1/2- \hat k_2)+(1-\hat\delta)\hat k_1(1/2- \hat k_1),\ \hat\sigma_{33}=\hat\delta\hat\beta\hat k_2(1/3-\hat\beta \hat k_2)+(1-\hat\delta)\hat k_1(1-\hat\alpha)\{1/3-\hat k_1(1-\hat\alpha)\},\ \hat\sigma_{12}=\hat\sigma_{21}=\hat\delta\hat k_2(\hat\beta-1/2)+(1-\hat\delta)\hat k_1(1/2-\hat\alpha),\ \hat\sigma_{13}=\hat\sigma_{31}=\hat\delta\hat k_2(\hat\beta^2/3-2\hat k_2/3)+(1-\hat\delta)\hat k_1\{(1-\hat\alpha)^2/3-2\hat k_1/3 \},\text{ and } \hat\sigma_{23}=\hat\sigma_{32}=\hat\delta\hat\beta\hat k_2(1/3-\hat k_2)+(1-\hat\delta)(1-\hat\alpha)\hat k_1(1/3- \hat k_1).$

%\bibliographystyle{rss}
%\bibliography{reference.bib}